\documentclass[11pt]{article}

\usepackage{amsmath,booktabs}
\usepackage{epsf,latexsym,amssymb,amsmath,amsthm,graphics}
\usepackage[mathscr]{eucal}
\usepackage{times}

\newtheorem{theorem}{Theorem}
\newtheorem{itlemma}{Lemma}[section]
\newtheorem{itproposition}[itlemma]{Proposition}
\newtheorem{itcorollary}[itlemma]{Corollary}
\newtheorem{itremark}[itlemma]{Remark}
\newtheorem{itremarks}[itlemma]{Remarks}
\newtheorem{itdefinition}[itlemma]{Definition}
\newtheorem{itexample}[itlemma]{Example}
\newtheorem{itfact}{Fact}[section]

\newenvironment{lemma}{\begin{itlemma}\rm}{\end{itlemma}} 

\newenvironment{fact}{\begin{itfact}\rm}{\end{itfact}} 
\newcommand{\Proof}{\noindent {\em Proof}.\ \ }

\newcommand{\ea}{ et. al.}
\newcommand{\step}{\theta}

\newcommand{\beqn}[1]{\begin{eqnarray}\label{#1}}
\newcommand{\eeqn}{\end{eqnarray}}
\newcommand{\beq}{\begin{eqnarray*}}
\newcommand{\eeq}{\end{eqnarray*}}
\newcommand{\bit}{\begin{itemize}}
\newcommand{\eit}{\end{itemize}}
\newcommand{\rf}[1]{~(\ref{#1})}

\newcommand{\halmos}{\rule{1ex}{1.4ex}}
\newcommand{\abs}[1]{|{#1}|}

\newcommand{\out}{h}
\newcommand{\px}{F}
\newcommand{\Y}{{\mathcal Y}} 
\newcommand{\N}{{\mathbb N}} 
\newcommand{\R}{{\mathbb R}} 
\newcommand{\Rpo}{{\mathbb R_{\geq0}}} 
\newcommand{\Rnpo}{{\mathbb R^n_{\geq0}}}

\newcommand{\wg}{\mbox{\it wg}}
\newcommand{\WG}{\mbox{WG}}
\newcommand{\IWG}{\mbox{IWG}}
\newcommand{\EWG}{\mbox{EWG}}
\newcommand{\en}{\mbox{\it en}}
\newcommand{\EN}{\mbox{EN}}
\newcommand{\hh}{\mbox{\it hh}}
\newcommand{\HH}{\mbox{HH}}
\newcommand{\ci}{\mbox{\it ci}}

\newcommand{\CI}{\mbox{CI}}
\newcommand{\CN}{\mbox{CN}}
\newcommand{\ptc}{\mbox{\it ptc}}
\newcommand{\PTC}{\mbox{PTC}}
\newcommand{\tT}{\mbox{\tiny T}}
\newcommand{\tB}{\mbox{\tiny B}}
\newcommand{\tWT}{\mbox{\tiny WT}}

\newcommand{\twg}{\mbox{\tiny\it wg}}
\newcommand{\tWG}{\mbox{\tiny WG}}
\newcommand{\tIWG}{\mbox{\tiny IWG}}
\newcommand{\tEWG}{\mbox{\tiny EWG}}
\newcommand{\ten}{\mbox{\tiny\it en}}
\newcommand{\tEN}{\mbox{\tiny EN}}
\newcommand{\thh}{\mbox{\tiny\it hh}}
\newcommand{\tHH}{\mbox{\tiny HH}}
\newcommand{\tci}{\mbox{\tiny\it ci}}

\newcommand{\tCI}{\mbox{\tiny CI}}
\newcommand{\tCN}{\mbox{\tiny CN}}
\newcommand{\tptc}{\mbox{\tiny\it ptc}}
\newcommand{\tPTC}{\mbox{\tiny PTC}}
\newcommand{\ren}{r_{\mbox{\tiny endo}}}
\newcommand{\rex}{r_{\mbox{\tiny exo}}}
\newcommand{\tfree}{r_{\mbox{\tiny free}}}

\newcommand{\GI}{G_{\mbox{\tiny C,I}}}
\newcommand{\GII}{G_{\mbox{\tiny C,II}}}
\newcommand{\GIII}{G_{\mbox{\tiny C,III}}}
\newcommand{\GIV}{G_{\mbox{\tiny C,IV}}}
\newcommand{\GA}{G_{\mbox{\tiny Auto}}}

\newcommand{\WTset}{\mathcal W}

\newcommand{\mut}{\mbox{\tiny mut}}

\renewcommand{\halmos}{\rule{1ex}{1.4ex}}
\renewcommand{\qed}{\hfill \halmos} 

\title{Shape, size and robustness: \\ feasible regions in the parameter space of biochemical networks}
\author{
Madalena Chaves\thanks{Project COMORE, INRIA Sophia Antipolis, 2004 Route des Lucioles - BP 93
        06902 Sophia Antipolis, France, {\tt\small mchaves@sophia.inria.fr}}, 
Eduardo D. Sontag\thanks{BioMaPS Institute for Quantitative Biology and Dep. of Mathematics,
        Rutgers University, Piscataway, NJ 08854 USA,
        {\tt\small sontag@math.rutgers.edu} } 
  and       
Anirvan M. Sengupta\thanks{BioMaPS Institute for Quantitative Biology and Dep. of Physics,  
        Rutgers University, Piscataway, NJ 08854 USA,
        {\tt\small anirvans@physics.rutgers.edu}} 
}

\date{}

\begin{document}

\maketitle

\begin{abstract}
The concept of robustness of regulatory networks has been closely related to 
the nature of the interactions among genes, and the capability of pattern maintenance 
or reproducibility. Defining this robustness property is a challenging task, but
mathematical models have often associated it to the volume of the space of admissible 
parameters. Not only the volume of the space but also its topology and geometry contain
information on essential aspects of the network, including feasible pathways, 
switching between two parallel pathways or distinct/disconnected active regions
of parameters. 
A general method is presented here to characterize the space of admissible parameters,
by writing it as a semi-algebraic set, and then theoretically analyzing its topology and
geometry, as well as volume. This method provides a more objective and complete measure 
of the robustness of a developmental module.
As an illustration, the segment polarity gene network is analyzed.
\end{abstract}

\section{Introduction}
For biological networks, the concept of robustness often expresses the idea that 
the system's regulatory functions should operate correctly under a variety of situations.
The network should respond appropriately to various stimulii and recognize meaningful 
ones (either harmful or favorable), but it should also ignore small (not meaningful) 
variations in the environment as well as
inescapable fluctuations in the abundances of biomolecules involved in the network ~\cite{asbl99,lsw99,dmmo00}. One might even speculate that if the networks malfunctions easily as a result of mutations then it has low chance of being selected by evolution. In that case one might expect a certain degree of mutational robustness \cite{dmmo00,sds02}.

\begin{figure}[t]
\centerline{
\scalebox{0.4}[0.4]{\includegraphics{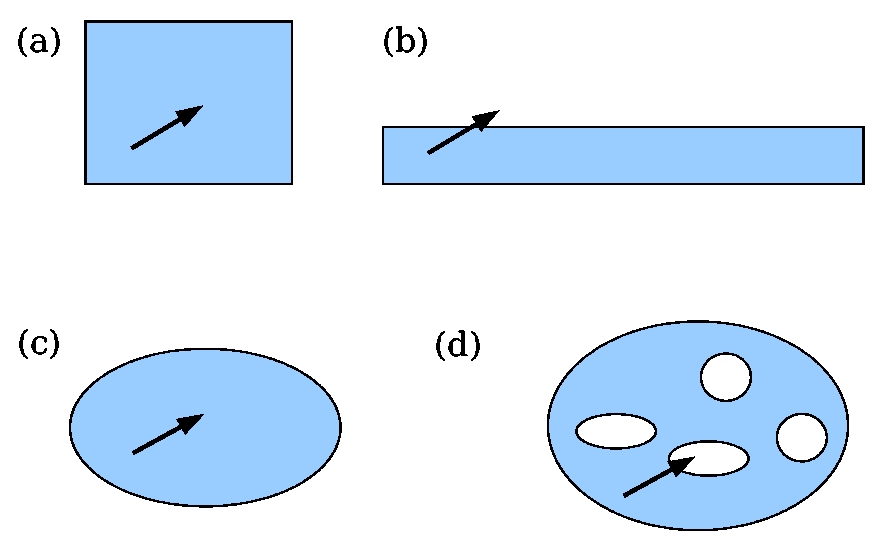} } }
\caption{The role of geometry and topology in robustness. Regions (a) and (b) have the same volume, 
but (b) is less robust: the same perturbation leads out of the space.
Regions (c) and (d) also have the same volume, but (d) is not a simply connected set, hence less robust.}
\label{fig-regions-shape}
\end{figure}

While it is difficult to define this robustness property in a precise form, it has been
associated to the space of admissible kinetic parameters, its volume~\cite{dmmo00}, 
and the effect of paramater perturbations on the qualitative behavior of the 
system~\cite{asbl99,lsw99}. Some methods for parameter sensitivity have been developed
~\cite{savageau71,heinrich96}, based essentially on derivatives of variables or fluxes with 
respect to the system's parameters.
The volume of the parameter space can be used as an indication of ``how many''
parameter combinations are possible, and these are related to the ability of the network
to work under a variety of situations. For instance, parameters may range through
different orders of magnitude, representing very different environments.
However, size is often not a reliable measure for robustness; other quantities, such as
shape, play a much more important role, as illustrated in Fig.~\ref{fig-regions-shape}.
Analysis of the shape or geometry of the admissible parameter set gives an indication not only
of its size, but also how far perturbations around each parameter disrupt the network.
A robust biological network will admit small fluctuations in its parameters without changing its
qualitative behavior. So, a robust network will be associated to a system whose parameter set
has few ``narrow pieces'' and ``sharp corners''. In such sets, reasonable parameter fluctuations 
may occur without leaving the set, hence maintaining the network's qualitative behavior
(compare Fig.~\ref{fig-regions-shape} (a) and (b)). 
We can formalize a measure of robustness that is related to having low rate of exit from the region under random walk \cite{sds02}. The rate of  first exit is intrinsically connected to the geometry of the region and is particularly sensitive to narrow directions and not just the overall volume.

To illustrate the importance of parameter space geometry, and the insight it brings to understanding 
the network, the model of the segment polarity network developed by von Dassow and 
collaborators~\cite{dmmo00} will be analyzed.
The segment polarity network is  part of a cascade of gene families responsible 
for generating the segmentation of the fruit fly embryo~\cite{sanson01}. 
Genes in earlier stages are transiently 
expressed, but the segment polarity genes maintain a stable pattern for about three hours. 
It has been suggested that the segment polarity genes constitute a robust developmental module, 
capable of autonomously reproducing the same behavior or generating the same gene expression 
pattern, in response to transient inputs~\cite{dmmo00,i04,ao03}. This robustness would be due to the
nature of interactions among genes, rather than the kinetic parameters of the reactions.
The model~\cite{dmmo00} describes the interactions among the principal segment polarity genes,
is continuous, and involves cell-to-cell
communications and around 50 parameters which are essentially unknown. The authors of~\cite{dmmo00} explored the 
model by randomly choosing 240,000 parameter sets out of which about 1,192 (or $0.5\%$) sets were 
consistent with the generation (at steady state) of the wild type pattern.
To explore the robustness of the network as a property of its interactions, Albert and Othmer~\cite{ao03}
developed a Boolean model of the segment polarity network, a discrete logical model where each 
species has only two states (0 or 1; ``OFF'' or ``ON''), but no kinetic parameters need to be defined.
This Boolean model is amenable to various methods for systematic robustness analysis~\cite{cas05,csa06,mlot06}. 
Ingolia~\cite{i04} focused on the properties of the (slightly changed) model~\cite{dmmo00} in individual 
cells, such as bistability, and extrapolated necessary conditions on parameters to the full intercellular 
model.

We propose a different approach, that retains the information contained on the kinetic parameters 
but reduces the model to a logical form with various possible ON levels and species-dependent
activation parameters. The admissible set of parameters of the model~\cite{dmmo00} 
is analyzed by constructing a cylindrical algebraic decomposition. 
Among other conclusions, our analysis completely explains the two ``missing links'' in von Dassow et. al.
original model, namely: why the segment polarity pattern can not be recovered without the negative 
regulation of {\em engrailed} by Cubitus repressor protein, and why the autocatalytic
{\em wingless} activation pathway vastly increases the network robustness. 
The present approach shows that, in contrast to volume only estimates, the topology and geometry of this set provide reliable quantitative measures of robustness of a system.

\section{Steady states define the feasible parameter space}

Previous studies~\cite{dmmo00,i04} have tested the parameter space
by randomly choosing sets of parameters and simulating the continuous model. 
If the corresponding trajectory reaches a steady state, and if this 
steady state is compatible with the experimentally observed wild type gene pattern, 
then the given set of parameters is said to be a ``solution'' to the modeling problem.

A more efficient and complete study of the parameter space can be devised, by first 
solving the algebraic equations of the model at steady state, and writing the steady state 
solutions as a function of the parameters. On the other hand, the steady state solutions 
are known -- 
the set of elements representing the wild type pattern is denoted by $\WTset$ 
-- so, one can then look for parameters 
that yield this pattern. Since many sets of parameters may be expected to yield the 
wild type pattern, this procedure provides a family of conditions defining regions 
of ``good''or feasible parameters ``$p$'' for wild-type steady states $x\in\WTset$.

\paragraph{The von Dassow\ea~model}
Before proceeding, recall that the model (Appendix~\ref{sec-original-eqs}) describes the 
concentrations of various mRNAs and proteins in a four cell parasegment of the fly embryo, 
subject to periodic boundary conditions (see also Fig.~\ref{fig-cells}). 
Here, each cell is assumed to have a square shape, with four faces (see Appendix~\ref{sec-WG}).
We next very briefly recall the species involved.
There are nine species with homogeneous concentration throughout each cell: engrailed mRNA 
and protein ($\en$ and $\EN$), wingless mRNA and (internal) protein ($\wg$ and $\IWG$), 
patched mRNA ($\ptc$), cubitus mRNA, active and repressor proteins ($\ci$, $\CI$, and $\CN$), 
and hedgehog mRNA ($\hh$).
Each of these species has a distinct concentration in each cell ($X_i$, $i=1,\ldots,4$).
In addition, there are three other species whose concentration varies in each of the four cell 
faces: external wingless protein ($\EWG$), patched protein ($\PTC$) and hedgehog protein ($\HH$).
For each of these species, the concentration in cell $i$ at face $j$ is denoted $X_{i,j}$,
$i=1,\ldots,4$, $j=1,\ldots,4$.
Thus, overall there are: $n=9\times4+3\times4\times4=84$ variables. Throughout the paper,
the following notation will be used (prime denotes transpose):
\beq
  X=(X_1,X_2,X_3,X_4)',  \mbox{ for } X\in\{\en,\EN,\wg,\IWG,\ptc,\ci,\CI,\CN,\hh\}.
\eeq
and
\beq
  X=(X_{1,1},X_{1,2},X_{1,3},X_{1,4},X_{2,1},\ldots,X_{4,4})',  
      \mbox{ for } X\in\{\EWG,\PTC,\HH\}.
\eeq
The total vector of concentrations is:
\beq
  x=(\en',\EN',\wg',\IWG',\EWG',\ptc',\PTC',\ci',\CI',\CN',\hh',\HH').
\eeq

\paragraph{Set of feasible parameters}
In general, the problem can be formulated mathematically by writing a set of equations 
dependent on the vector of species concentrations ($x\in\Rnpo$) and the parameter vector ($p\in\R^r_{\geq0}$), 
together with a set of outputs ($y\in\R^m_{\geq0}$, the available gene expression levels).
Introduce functions $f:\Rnpo\times\R^r_{\geq0}\to\R^n$ and $\out:\Rnpo\to\Y\subset\R^m_{\geq0}$,
where $\Rpo=\{x\in\R: x_i\geq0, \mbox{ for all $i$} \}$, and consider the system
with outputs
\beqn{eq-sys}
  \frac{dx}{dt} &=& f(x,p)\\
     y &=& \out(x)
\eeqn
where the function $h(x)$ could be, for instance, a vector listing the concentration of 
{\em wingless}, {\em engrailed}, {\em hedgehog} and {\em cubitus}, four of the
segment polarity mRNAs which have been experimentally measured. 
Or, in other words, $y$ is ``the phenotype corresponding to the genotype $x$''.
The wild-type gene expression output set can be defined as:
\beq
  \Y^{\tWT}=\{y\in\Y:\ y=\out(x),\ x\in\WTset\}.
\eeq
The problem of characterizing the sets of feasible parameters is then reduced to finding all possible 
parameter vectors $p$ which lead the system to have an output in $\Y^{\tWT}$, at steady state. 
This will be the set of ``good'' parameters:
\beqn{eq-good-set}
    G=\{\ p\in\R^r_{\geq0}:\ \exists x \mbox{ s.t. } f(x,p)=0 \mbox{ and } \out(x)\in\Y^{\tWT}\ \}.
\eeqn

\paragraph{Large Hill coefficients}
A straightforward approach would be to solve the original system at steady state,
obtain expressions for $x\in\WTset$ in terms of $p$, and compare these expressions to the outputs 
in $\Y^{\tWT}$:
\beq
   f(x,p)=0 \ \Leftrightarrow\ x =F(p) 
  \ \ \ \mbox{and}\ \ \ F(p)\in \WTset  
   \Leftrightarrow\ p\in G .
\eeq
A possible drawback of this method is that explicit solutions $x=F(p)$ for the original system 
and then explicit formulas for $G$ may not be easy to compute. 
On the other hand, many of the equations in the model~\cite{dmmo00} involve terms of the 
form (see also Appendix~\ref{sec-original-eqs}):
\beq
  \phi(X,\kappa,\nu) = \frac{X^\nu}{\kappa^\nu+X^\nu},
\eeq
meaning that the function $\phi$ is active (ON), if species $X$ is above a certain 
threshold $\kappa$. The exponent $\nu$, also known as the Hill coefficient, characterizes the steepness of an OFF/ON transition. 
For large enough exponents, this saturation function becomes very steep,
and $\phi$ becomes practically insensitive to the actual value of $\nu$. 
As found in~\cite{do02}, coefficients $\nu$ must indeed be quite large for the network to achieve 
robustness: namely in the interval $[5.0,10.0]$. 
This is also the basis of the typical on/off logical interpretation of gene expression.
Any such term $\phi(X,\kappa,\nu)$, for large $\nu$, may thus be replaced by a step 
function with two levels (0 or 1):
 \beq
   \step(X-\kappa)= 
     \left\{
       \begin{array}{ll}
          0, & X <\kappa \\
          1, & X >\kappa\ . \\
       \end{array}
    \right.
\eeq
Thus, when $\nu$ is large:
\beqn{eq-phi-step}
  \lim_{\nu\to\infty}\,\phi(X,\kappa,\nu) = \step(X-\kappa),\ \ \ 
  \lim_{\nu\to\infty}\,\psi(X,\kappa,\nu) = 1-\step(X-\kappa) = \step(\kappa-X).
\eeqn
A composite function of $\phi$ and $\psi$ also frequently appears in the continuous equations 
(Appendix~\ref{sec-original-eqs}):
\beq
   \phi(X_a\psi(X_b\kappa_b,\nu_b),\kappa_a,\nu_a).
\eeq
This function can be simplified in terms of step functions to:
\beq
   \step(X_a\step(\kappa_b-X_b)-\kappa_a)=
   \step(X_a-\kappa_a)\step(\kappa_b-X_b)
\eeq
since
\beq
 &&  X_b>\kappa_b \Rightarrow \step(\kappa_b-X_b)=0 
   \Rightarrow \step(X_a\step(\kappa_b-X_b)-\kappa_a)=\step(-\kappa_a)=0,
  \\
 &&  X_b<\kappa_b \Rightarrow \step(\kappa_b-X_b)=1
   \Rightarrow \step(X_a\step(\kappa_b-X_b)-\kappa_a)=\step(X_a-\kappa_a).
\eeq
As an example,  consider the equation governing {\em engrailed}
from the original model which can be found in~\cite{dmmo00,do02}
(or in Appendix~\ref{sec-original-eqs}). 
In this model the concentration of {\em engrailed} in cell $i$ ($\en_i$), is positively regulated 
by external Wingless protein ($\EWG_{\underline i}$) and negatively regulated by Cubitus repressor 
protein ($\CN_i$) concentrations (further notation is found in Appendix~\ref{sec-notation}):
\beq
  \frac{d\en_i}{dt}&=&\frac{1}{H_{\ten}}
                      \left( 
                        \phi(\EWG_{\underline i}\psi(\CN_i,\kappa_{\tCN\ten},\nu_{\tCN\ten}),
                               \kappa_{\tWG\ten},\nu_{\tWG\ten})
                        -\en_i 
                      \right) .
\eeq
For large exponents $\nu$, this simplifies to the equation:
\beq
  \frac{d\en_i}{dt}&=&\frac{1}{H_{\ten}}
                      \left( 
                         \step(\EWG_{\underline i}-\kappa_{\tWG\ten})
                         \step(\kappa_{\tCN\ten}-\CN_i)
                         -\en_i 
                      \right). 
\eeq
To analyticaly study the space of feasible parameters for the segment polarity network 
model~\cite{dmmo00}, we will thus consider that all exponents $\nu$ are large, and apply 
method\rf{eq-phi-step} to simplify the original system of equations. 
The von Dassow\ea~model is then characterized by equations\rf{eq-d-en}-(\ref{eq-d-HH})
(Appendix~\ref{sec-discrete-syst}).
The parameters are as in~\cite{dmmo00}, except $T_i$ and $U_i$, which represent
the maximal values of $\ptc$ and $\ci$ (respectively), in each cell. 
These take values in the interval $[0,1]$ and generalize the possible ON values of 
$\ptc$ and $\ci$ (to be discussed later).
In addition, as discussed, the system is assumed to be at steady state, in which case 
the gene expression pattern must satisfy:
\beq
  \en_i = \step(\EWG_{\underline i}-\kappa_{\tWG\ten})\step(\kappa_{\tCN\ten}-\CN_i).
\eeq
Applying\rf{eq-phi-step} and then solving the system at steady state yields the set of 
algebraic equations\rf{eq-discrete-en}-(\ref{eq-discrete-HH}), which characterize the gene expression 
pattern of the segment polarity network according to the von Dassow\ea~model.

\paragraph{Maximal (ON) expression levels}
While some of the species have a normalized maximal expression level (to 1), such as $\en$ or $\hh$,
other species may be more generally allowed to have any positive value (namely, $\ptc$ and $\ci$). 
These maximal expression levels are also treated as parameters.
When using\rf{eq-phi-step} to simplify the {\em patched} equation\rf{eq-c-ptc} to\rf{eq-d-ptc}, 
we have generalized the equation and added distinct maximal levels of expression in each cell, 
given by $T_i$ ($i=1,\ldots,4$). This allows a more accurate representation of experimental data,
which shows that {\em patched} is strongly expressed in every second and fourth cells,
weakly expressed in every first cell, and not expressed in every third cell 
(see~\cite{dmmo00} for more discussion). Thus we will consider $T_1<T_2=T_4$: 
\beqn{eq-ptc}
   \ptc_{1,2,3,4}^{\tWT}=(T_1,T_2,0,T_2)'.
\eeqn
A similar generalization was made to deal with the activation of {\em cubitus interruptus}.
In von Dassow\ea~model, this is due to some external parameters $B_i$ (not governed by
a dynamical equation), with a corresponding activity threshold $\kappa_{\tB\tci}$. 
However, for more generality, and to allow  distinct maximal levels of expression in each cell,
we have replaced each of the terms $\step(B_i-\kappa_{\tB\tci})$ in\rf{eq-d-ci} by a parameter
$U_i$, $i=1,\ldots,4$\rf{eq-discrete-ci}.
Furthermore, in characterizing the set of feasible parameters, it will become clear that
allowing distinct $U_i$ enlarges the space of possible parameters, by introducing the four 
regions $\GI$ to $\GIV$.
Thus the steady state values for the cubitus mRNA are:
\beqn{eq-ci}
   \ci_{1,2,3,4}^{\tWT}=(U_1,U_2,0,U_4)'.
\eeqn
Asymmetry in {\em cubitus} expression (i.e., distinct values $U_i$) 
could be due, for instance, to some of the pair rule genes. Sloppy paired, or 
a combination of Runt and Factor X, regulate the transition from pair rule to segment polarity genes
expression, and induce asymmetric anterior/posterior parasegment expression~\cite{sg04}. 

Finally, note that the maximal expression levels of $\wg$ are expressed in terms of the parameters 
$\alpha_{\tCI\twg}$ and $\alpha_{\tWG\twg}$. 
From equation\rf{eq-discrete-wg}, there are three possible combinations of the step functions,
each leading to a different value for $\wg_2$. These three possibilities are:
\beq
  w_{\tCI}=\frac{\alpha_{\tCI\twg}}{1+\alpha_{\tCI\twg}},\ \
  w_{\tCI,\tWG}=\frac{\alpha_{\tCI\twg}+\alpha_{\tWG\twg}}{1+\alpha_{\tCI\twg}+\alpha_{\tWG\twg}},\ \
  w_{\tWG}=\frac{\alpha_{\tWG\twg}}{1+\alpha_{\tWG\twg}},
\eeq
and each reflects a different pathway for {\it wingless} activation. Indeed, {\it wingless} 
can be activated by Cubitus only (in which case the maximal amplitude is given by $w_{\tCI}$), by both 
Cubitus and Wingless ($w_{\tCI,\tWG}$), or by Wingless only ($w_{\tWG}$).

\paragraph{Outputs}
The next question concerns the choice of an appropriate output function.
The gene expression patterns for {\em engrailed}, {\em wingless}, {\em hedgehog}, {\em cubitus}, 
and {\em patched} are among the most well documented, so we will consider the output function 
$\out:\Rnpo\to\R_{\geq0}^{20}$:
\beqn{eq-output}
   y=\out(x)=\begin{pmatrix} \out_{\ten}(x)\\ \out_{\twg}(x)\\ 
                    \out_{\tptc}(x)\\ \out_{\tci}(x)\\ \out_{\thh}(x) \end{pmatrix}.
\eeqn
At steady state, both $\en$ and $\hh$ are expressed in every third cell~\cite{hi90}, 
which translates into
\beqn{eq-en-hh}
   \out_{\ten}(x)=(0,0,1,0)',\ \ \ \out_{\thh}(x)=(0,0,1,0)',
  \ \ \mbox{ for } x\in\WTset.
\eeqn
Further experimental observations show that {\it cubitus} is expressed in all but the 
third cell~\cite{ek90}, and {\it patched} is strongly expressed in every second and fourth~\cite{hi90},
but more weakly expressed in every first cell. So:
\beqn{eq-ci-ptc}
   \out_{\tci}(x)=(U_1,U_2,0,U_4)',\ \ \ \out_{\tptc}(x)=(T_1,T_2,0,T_2)',
  \ \ \mbox{ for } x\in\WTset.
\eeqn
Finally,  {\it wingless} ($\wg$) is only expressed in every second cell~\cite{hi90}, 
to the left of $\en$, that is: 
\beqn{eq-wt1}
   \out_{\twg}(x)=(0,w,0,0)',\ \ \mbox{ for } x\in\WTset.
\eeqn
To summarize, in this example, the set of output values at steady state is: 
\beqn{eq-Y_WT}
   \Y^{\tWT}=\left\{((0,0,1,0),(0,w,0,0),(T_1,T_2,0,T_2),(U_1,U_2,0,U_4),(0,0,1,0))': \right. \nonumber\\
                \left. w,T_1,T_2,U_1,U_2,U_4>0,\ T_1<T_2\right\}. 
\eeqn
The first result to be noted is that there is a {\em unique} steady state $x=x(p)\in\WTset$ for
each set of parameters $p$: 
\begin{theorem}\label{th-unique-ss}
Let $f$ be the function $\Rnpo\times\R^r_{\geq0}\to\R^n$ given by\rf{eq-d-en}-(\ref{eq-d-HH}),
and $\out$ be the function $\Rnpo\to\Y$ given by\rf{eq-output}. Define $G$ as in\rf{eq-good-set}.
Then, there exists a function $\px:G\to\Rnpo$ such that, for each $p\in\R^r_{\geq0}$ and each $x\in\Rnpo$,
\beq
   f(x,p)=0\ \ \mbox{and} \ \ \out(x)\in\Y^{\tWT}  
\ \ \mbox{imply}\ \ x=\px(p).
\eeq
\end{theorem}

\Proof
Pick any $p\in G$, and an $x\in\Rnpo$ satisfying $f(x,p)=0$ and $\out(x)\in\Y^{\tWT}$.
The equations $f(x,p)=0$ can be simplified to yield\rf{eq-discrete-en}-(\ref{eq-discrete-HH}).
We must check that these equations are all consistent and admit only one solution.
Since all mRNAs $\en$, $\wg$, $\ptc$, $\ci$, and $\hh$ are provided by $\out(x)$, we must solve 
for the proteins, and then substitute these back into the equations for the mRNAs, to check consistency.

The Engrailed protein is straigthforward: $\EN=\en$.
We start by solving for $\EWG$, with $\wg=(0,w,0,0)'$ as given.
First note that the matrix $M$ is diagonally dominant, by adding up the entries in any column:
\beqn{eq-sumcol}
   -\left(H_{\tIWG}^{-1}+\ren+r_{M}+2r_{LM}\right) 
   +2r_{LM}+r_M+4h = -H_{\tIWG}^{-1} -\ren\,\frac{1}{1+H_{\tIWG}\rex}
\eeqn
which is always a negative quantity. By Ger\v sgorin's Theorem, all eigenvalues of $M$ are 
contained in the disk centered at $-d+h$ with radius $2r_{LM}+r_M+3h$, therefore all eigenvalues
have negative real parts. 
Thus, the matrix $M$ is symmetric and negative definite, and since the right-hand-side vector 
in \rf{eq-discrete-EWG} is also non-positive, there is a unique solution 
\beq
   \EWG=-\frac{1}{4}\frac{\rex}{1+H_{\tIWG}\rex}\, M^{-1}\ \widetilde{\wg}
\eeq
which is real and positive, {\it for each set of parameters $p$}.
Once we have $\EWG$, we can immediately solve for $\IWG$ from\rf{eq-discrete-IWG}.

The solution for $\PTC$ and $\HH$ can also be exactly and uniquely computed from\rf{eq-discrete-PTC}
and\rf{eq-discrete-HH}, for any output $\ptc=(T_1,T_2,0,T_2)'$ (this calculation is shown
Appendix~\ref{sec-PTC-HH}).

Finally, one can now straightforwardly and uniquely compute the values of $\CI$ and $\CN$, 
from\rf{eq-discrete-CI} and\rf{eq-discrete-CN}, and the values of $\ci$ and $\PTC$.

The last step is the substitution of $\EN$, $\EWG$, $\IWG$, $\PTC$, $\HH$, $\CI$ and $\CN$ back into the
equations for the mRNAs\rf{eq-discrete-en},\rf{eq-discrete-wg},\rf{eq-discrete-ptc},\rf{eq-discrete-ci},
and\rf{eq-discrete-hh}. But, since $p\in G$, by definition we are guaranteed that these
equalities are indeed satisfied. 
\qed

\paragraph{Missing link: {\em engrailed} regulation by Cubitus repressor}
A second result from our model formulation is the explanation of a ``missing link''
in a first version of the model proposed by von Dassow\ea~\cite{dmmo00}.
In this first version, {\em engrailed} was regulated only by $\EWG$, and no feasible parameter sets
were found. 
Indeed, below (Theorem~\ref{th-EWG-symmetry}) we prove that, {\em for any set of parameters},
the mechanism for {\it wingless} regulation generates a strong symmetry in the steady state expression 
of external Wingless. This symmetry effectively prevents any asymmetry arising in $\en$ due to $\EWG$ only.
\begin{theorem}\label{th-EWG-symmetry}
Let $w>0$ and assume $\wg^{\tWT}=(0,w,0,0)'$. Then, at steady state:
\beqn{eq-EWG-symmetry}
    \EWG_{\underline 4}^{\tWT}<\EWG_{\underline 1}^{\tWT}
                              =\EWG_{\underline 3}^{\tWT}<\EWG_{\underline 2}^{\tWT}.
\eeqn 
\qed
\end{theorem}
The proof is based on a sequence of algebraic calculations, and is shown in Appendix~\ref{sec-WG}.
Now, consider the steady state equation for {\em engrailed}, when no dependence on $\CN$ is assumed:
\beq
   \en_i^{\tWT}=\step(\EWG_{\underline i}^{\tWT}-\kappa_{\tWG\ten})
\eeq
Compare to the output\rf{eq-en-hh}:
\beq
   \out_{\ten}(x)=(0,0,1,0)'.
\eeq
Then, from the definition of $\step$, for consistency in our model it is necessary that:
\beq
  && \EWG_{\underline i}^{\tWT}<\kappa_{\tWG\ten},\ \ \mbox{ for } i=1,2,4 \\
  && \EWG_{\underline 3}^{\tWT}>\kappa_{\tWG\ten}.
\eeq
However, by\rf{eq-EWG-symmetry}, the inequalities for
$i=1,2$ and $i=3$ are incompatible. This means that, due to the symmetry in Wingless distribution, 
such a simple regulation of $\en$ {\em can never lead to the segment polarity pattern}.
Thus {\it engrailed} requires regulation by some other factor, in this case repression by
the Cubitus protein ($\CN$), as in\rf{eq-discrete-en}.
In order to obtain repression of $\en$ in the first and second cells, one can now ask:
\beq
  && \CN_1^{\tWT}>\kappa_{\tCN\ten},\ \ \CN_2^{\tWT}>\kappa_{\tCN\ten}\\
  && \EWG_{\underline 3}^{\tWT}>\kappa_{\tWG\ten} \mbox{ and } 
     \CN_3^{\tWT}<\kappa_{\tCN\ten}\\
  && \EWG_{\underline 4}^{\tWT}<\kappa_{\tWG\ten} \mbox{ or } 
     \CN_4^{\tWT}>\kappa_{\tCN\ten}
\eeq
that is, $\CN$ is responsible for repression in both the first and second cells.
This means that, at steady state, $\CN$ must be expressed in both the first and second cells.
This in turn requires the presence of Patched protein in both the first and second cells. 
On the other hand, from Appendix~\ref{sec-PTC-HH}, we know that a steady state $x\in\WTset$ with
$\out(x)\in\Y^{\tWT}$, implies $\ptc_1^{\tWT}=\PTC_1^{\tWT}$, and also 
$\PTC_2^{\tWT}=\PTC_4^{\tWT}$.
This can be stated as:
\begin{lemma}\label{lm-ptc1}
Consider system\rf{eq-sys} and assume that, at steady state, the output set is  $\Y^{\tWT}$.
Then $\ptc_1^{\tWT}=\PTC_1^{\tWT}\neq0$. If $\ptc^{\tWT}=(T_1,T_2,0,T_2)'$ with
$T_1<T_2$, then $\PTC_1^{\tWT}=T_1$ and $\PTC_2^{\tWT}=\PTC_4^{\tWT}>0$.
\qed
\end{lemma}
While {\em patched} expression is typically weaker in the first than in second and fourth cells
(see~\cite{dmmo00}), this shows that it is nevertheless necessary, that is, the segment polarity
gene pattern obtains only when $T_1>0$.
The discussion on $\CN$ leads to the following conclusion:
\begin{lemma}\label{lm-CI-CN}
Consider system\rf{eq-sys} and assume that, at steady state, the output set is  $\Y^{\tWT}$.
Let $\ptc^{\tWT}=(T_1,T_2,0,T_2)'$ with $T_1<T_2$.
Then $\PTC_{1,2}^{\tWT}>\kappa_{\tPTC\tCI}$ and 
\beqn{eq-CI-CN}
   \CI_i^{\tWT} = U_i\,\frac{1}{1+H_{\tCI}C_{\tCI}},\ \ \ 
   \CN_i^{\tWT} = U_i\,\frac{H_{\tCI}C_{\tCI}}{1+H_{\tCI}C_{\tCI}},\ \ \ i=1,2,4,
\eeqn
and $\CI_3^{\tWT}=\CN_3^{\tWT}=\ci_3^{\tWT}=0$.
\end{lemma}

\Proof
Theorem~\ref{th-EWG-symmetry} and the subsequent discussion shows that $\CN_{1,2}^{\tWT}\neq0$ is needed.
From\rf{eq-discrete-CN}, this can only be achieved by asking $\PTC_{1,2}^{\tWT}>\kappa_{\tPTC\tCI}$. 
By Lemma~\ref{lm-ptc1}, it also holds that $\PTC_2^{\tWT}=\PTC_4^{\tWT}>\kappa_{\tPTC\tCI}$.
This means that both\rf{eq-discrete-CI} and\rf{eq-discrete-CN} can be simplified to\rf{eq-CI-CN},
at steady state. On the third cell, $\CI_3^{\tWT}=\CN_3^{\tWT}=0$ because the output is zero.
\qed

\section{A cylindrical algebraic decomposition of the parameter space}
The algebraic equations $f(x,p)=0$ together with $\out(x)\in\Y^{\tWT}$ are a representation of the set of 
good parameters $G$, though not providing as yet explicit conditions on $p$.
An explicit characterization of the parameters $p$ may be obtained by calculating a
cylindrical algebraic decomposition (CAD) of $G$: this is a special type of representation of $G$ 
as a finite union of disjoint connected components. 
A CAD will provide a hierarchy of inequalities on $p_1$, $p_2$,$\ldots$, $p_r$, from 
which the volume of $G$, as well as its geometry and topology, may be deduced.

Computing the cylindrical algebraic decomposition of a semi-algebraic set is a 
complex problem, but various standard algorithms are available~\cite{c75,acm84}. 
Several software packages have been developed, for instance QEPAD~\cite{qepad}, 
(based in~\cite{ch91}) and in Mathematica~\cite{wolfram}. See also~\cite{nmgj01} for an 
overview of available software, current applications, and many other related references.  
Common applications of CADs include computation of the controllable or reachabable sets
in hybrid systems~\cite{gt04}.
Constructing a CAD involves the use of symbolic computation and, while various improvements
have been achieved, it still is a time consuming problem. 
For instance, the estimated maximum time for the algorithm~\cite{c75} is dominated by 
``$2^{2^{kN}}$'', where $N$ is the length of the input formula and $0<k\leq8$.
Fortunately, in view of these computational complexity difficulties, in the present example it 
is relatively easy to directly compute a CAD without using general methods, and we will do so.

For equations~\rf{eq-discrete-en}-(\ref{eq-discrete-HH}), subject to\rf{eq-en-hh}-(\ref{eq-wt1}), 
a cylindrical algebraic decomposition can be constructed in which several parameters 
(Table~\ref{par-free}) are free to take any values (within physiological restrictions only).
At the next level, parameters in Table~\ref{par-common} have constraints which depend
only on those parameters given in Table~\ref{par-free}. The last level is formed by the 
parameters in Table~\ref{par-regions}, whose constraints depend on parameters from both 
previous levels (Tables~\ref{par-free} and~\ref{par-common}),
thus defining a polyhedron.

Following the model of von Dassow et. al., there are two possible parallel pathways for {\it wingless} 
activation: either by the Cubitus interruptus protein ($\CI$), or through auto-activation; 
both pathways could be simultaneously activating {\it wingless} production.
Since the activation constants $\alpha_{\tCI\twg}$ and  $\alpha_{\tWG\twg}$, 
are free parameters, in each of the three cases $\wg_2^{\tWT}$ will have
a different ON level (respectively, $w_{\tCI}$, $w_{\tWG}$, or $w_{\tCI,\tWG}$).
Computation of  $\EWG$ and $\IWG$ depends on $\wg_2^{\tWT}$, so each of these three
cases must be separately analyzed for feasibility.
For both pathways, exact analytic computation of $\PTC_{i,j}$ and $\HH_{i,j}$ ($i,j=1,\ldots,4$)
is also carried out (see Appendix~\ref{sec-PTC-HH}). 
Several disconnected regions of parameters will be defined by the levels of {\it cubitus}, $U_{1,2,4}$.

\paragraph{Five disconnected regions}
When only $\CI$ and $\CN$ regulate {\it wingless} expression, it is easy to 
see from\rf{eq-discrete-wg},\rf{eq-wt1} and\rf{eq-CI-CN}  that:
\beqn{eq-wgU134}
  && \left(
     U_i\frac{1}{1+H_{\tCI}C_{\tCI}} < \kappa_{\tCI\twg} \ \ \mbox{ or }\ \ 
     U_i\frac{H_{\tCI}C_{\tCI}}{1+H_{\tCI}C_{\tCI}}> \kappa_{\tCN\twg} 
     \right)
     \ \ \mbox{ and }\ \ \IWG_i < \kappa_{\tWG\twg}
\eeqn
for $i=1,3,4$, and
\beqn{eq-wgU2}
   U_2\frac{1}{1+H_{\tCI}C_{\tCI}} > \kappa_{\tCI\twg} \ \ \mbox{ and }\ \ 
     U_2\frac{H_{\tCI}C_{\tCI}}{1+H_{\tCI}C_{\tCI}}< \kappa_{\tCN\twg}
     \ \ \mbox{ and }\ \ \IWG_2 < \kappa_{\tWG\twg}\ .
\eeqn
From observation of\rf{eq-wgU134},\rf{eq-wgU2} it is clear that the situations
$U_2=U_1$ or $U_2=U_4$ are not well defined, since contradictory constraints 
are imposed on $\kappa_{\tCI\twg}$ and  $\kappa_{\tCN\twg}$. So, the
regions of parameters satisfying  $U_2=U_1$ or $U_2=U_4$ are not feasible.
This divides the set $G$ into at least four disconnected components, divided
by the hyperplanes $U_2=U_1$ or $U_2=U_4$ ($\GI$, $\GII$, $\GIII$, and $\GIV$ in Fig.~\ref{fig-cubeU124}).
A similar argument holds for the case when both pathways
contribute to activation of {\it wingless} on the second cell.
The four disconnected regions of parameters are identified in Table~\ref{par-regions}. 

Finally, the third case (auto-activation pathway only), introduces a fifth component 
of $G$ ($\GA$), which must be disconnected from either of the previous four components. 
This is clear, by contrasting the necessary conditions in the second cell for
either case (compare to\rf{eq-wgU2}):
\beqn{eq-wgU2E}
    \left(
    U_2\frac{1}{1+H_{\tCI}C_{\tCI}} < \kappa_{\tCI\twg} \ \ \mbox{ or }\ \ 
    U_2\frac{H_{\tCI}C_{\tCI}}{1+H_{\tCI}C_{\tCI}}> \kappa_{\tCN\twg} 
    \right)
     \ \ \mbox{ and }\ \ \IWG_2 > \kappa_{\tWG\twg}.
\eeqn
The five disconnected components are thus first defined by $U_1$, $U_2$, and $U_4$, and then by
$\kappa_{\tCI\twg}$ and $\kappa_{\tCN\twg}$. 
The projection on the ($\kappa_{\tCI\twg}$,$\kappa_{\tCN\twg}$,$\kappa_{\tWG\twg}$)-dimensions
compares two of these components ($\GII$ and $\GA$), both polyhedrons (Fig.~\ref{fig-cube-kwg}).

The cylindrical algebraic decomposition is shown in detail in Appendix~\ref{sec-CAD}, and
summarized in Tables~\ref{par-free},\ref{par-common},\ref{par-regions}. 
Each of the five components, $G_{\gamma}\in\{\GI,\GII,\GIII,\GIV,\GA\}$  
is thus described by a hierarchy of sets of the form:
\beqn{eq-SN}
   S_{1,\gamma} &=& (a_{\gamma},b_{\gamma})\subset\R \nonumber\\
   S_{i,\gamma} &=& \{(x,x_i)\in\R^i:\ x\in S_{i-1,\gamma},\ 
                       \alpha_{i,\gamma}(x)<x_i<\beta_{i,\gamma}(x)\} \subset \R^i 
\eeqn
for $i=2,\ldots,N$, where $\alpha_{i,\gamma},\beta_{i,\gamma}: S_{i-1,\gamma}\to\R_{>0}$ and 
$S_{N,\gamma}=G_{\gamma}$. 
It can be shown that each $G_{\gamma}$ is in fact topologically equivalent to the
unitary open hypercube, and hence topologically trivial.

\begin{theorem}
For each $G_{\gamma}$, the set $S_N=S_{N,\gamma}$, as obtained from\rf{eq-SN}, is homeomorphic to $(0,1)^N$. 
\end{theorem}

\Proof
Pick any $G_{\gamma}$, and drop the subscript $\gamma$, for simplicity of notation.
To argue by induction, note that the set $S_1$ is clearly homeomorphic to $(0,1)$. 
For $i\geq2$, assume that $S_{i-1}$ is homeomorphic to $(0,1)^{i-1}$.  
Next, define the following continuous function:
\beq
  \varphi_i: S_{i-1}\times(0,1)\to S_{i-1}\times\R, \ \ \
   \varphi_i(x,t)=(x,f_i(x)+t\,(g_i(x)-f_i(x))).
\eeq
For each fixed $x$, $\alpha_i(x)<\alpha_i(x)+t\,(\beta_i(x)-\alpha_i(x))<\beta_i(x)$ for all $t\in(0,1)$.
Therefore, $\varphi_i$ maps into $S_i$.
On the other hand, $\varphi_i$ has an inverse function defined on $S_i$ and continuous, given by:
\beq
   \varphi_i^{-1}: S_{i}\to S_{i-1}\times(0,1),\ \ \
   \varphi_i^{-1}(x,y)=\left(x,\frac{y-\alpha_i(x)}{\beta_i(x)-\alpha_i(x)}\right).
\eeq
So $S_i$ is homeomorphic to $S_{i-1}\times(0,1)$, and therefore, by inductive
hypothesis, to $(0,1)^i$, as we wanted to show.\qed

\paragraph{Relative volume and the second missing link}
Once the parameter set $G$ is characterized by writing intervals for the 
various parameters in the form \rf{eq-SN}, it is very easy to compute the (relative) volumes of the disconnected components. 
Note that in each component only the intervals for $U_1$, $U_2$, $U_4$, and 
$\kappa_{\tCN\ten}$, $\kappa_{\tWG\ten}$, 
$\kappa_{\tCI\twg}$, $\kappa_{\tCN\twg}$, $\kappa_{\tWG\twg}$ vary; constraints on the remaining
parameters are common to all components.
Following a Monte Carlo approach, the parameters in Tables~\ref{par-free},\ref{par-common} are chosen first, 
and then $\kappa_{\tCI\twg}$, $\kappa_{\tCN\twg}$, $\kappa_{\tWG\twg}$ from the unitary cube
(all parameters are randomly chosen from uniform distributions in the given intervals).
It is next checked whether the parameter set falls in any of the components $\GI$, $\GII$, 
$\GIII$, $\GIV$, $\GA$, or outside $G$.
This method provides an estimate of the volumes of each disconnected component, when projected
into the 
($\kappa_{\tCN\ten}$,$\kappa_{\tWG\ten}$,$\kappa_{\tCI\twg}$,$\kappa_{\tCN\twg}$,$\kappa_{\tWG\twg}$)
dimensions, as the fraction of parameter sets that fall into each component.
The volume of this 5-dimensional cube occupied by feasible parameter sets is only about 0.7\%.
As is illustrated by the polyhedrons in Fig.~\ref{fig-cube-kwg}, component $\GA$ is much larger than 
the others -- approximately 40 to 270 times larger.
\begin{table}[ht]
\caption{Relative volumes of the five disconnected components. 
In component $\GA$, only auto-activation leads to {\it wingless} expression. 
In components $\GI$, $\GII$, $\GIII$, and $\GIV$, $\CI$ always activates 
{\it wingless} expression. Total number of parameter sets generated: $1\times10^7$.
Number of feasible parameter sets: 70026.}
\label{tb-volume}
\begin{center}
\begin{tabular}{cc}
\hline
  Component    & Volume  \\
\hline
  $\GI$   &  $1.6\times10^{-4}$  \\
  $\GII$   &  $0.25\times10^{-4}$  \\
  $\GIII$   &  $0.86\times10^{-4}$  \\
  $\GIV$   &  $0.46\times10^{-4}$  \\
  $\GA$   &  $67\times10^{-4}$  \\
\hline
\end{tabular} 
\end{center}
\end{table}

The large difference observed between $\GI$-$\GIV$ and $\GA$ explains the second
``missing link'' in the first version of von Dassow\ea~model, namely the {\em wingless} 
autocatalytic activation. Note that the presence of this link greatly increases
the total volume of the feasible parameter space: in fact the  region $\GA$ is
95\% of the total volume.

\paragraph{Parameter tendencies}
As described above, the parameter space for the segment polarity network can be described by a CAD, 
a hierarchy of inequalities on the parameters where an interval is explicitly given for each parameter.
At the base of this hierarchy, there is a first group of parameters whose intervals correspond
simply to physiological values, as in Table~\ref{par-free}. 
The intervals for the remaining parameters have bounds which depend on the parameters in the 
first group (Tables~\ref{par-common} and~\ref{par-regions}).
In any case, one may ask how the parameters are distributed in their intervals,
whether each parameter $p_i$ is more likely to attain high or low values more frequently, 
or whether a ``tendency'' for each parameter $p_i$ be identified.
An answer to this question is obtained by randomly generating parameters in the full 
parameter space $G$, and computing the distribution of each parameter. Taking all the parameter sets
generated to compute the relative volumes of the  five disconnected components of $G$, and computing
a histogram for each parameter, the result shown on Fig.~\ref{fig-par-dist} is obtained. 
As expected, many parameters have a uniform distribution, as their values do not influence the final 
outcome of the network in any particular way (for instance, most half-lives).
Other parameters  exhibit a marked tendency for higher (e.g., $\kappa_{\tCN\tptc}$), medium 
(e.g., $\kappa_{\tWG\twg}$) or lower (e.g., $\kappa_{\tCI\tptc}$) values. 
All the parameters that exhibit a marked tendency are listed in Table~\ref{tb-par-space}, 
and classified according to their function in the network: for instance,  
$\kappa_{\tCN\tptc}$ represents the repression of $\ptc$ by $\CN$,
and therefore, high values of $\kappa_{\tCN\tptc}$ correspond to a weak repression.

A very similar analysis was performed by von Dassow and Odell~\cite{do02}, who also plotted 
the distribution of their family of feasible parameters to determine possible constraints for 
each parameter. 
Overall, our results agree very well with those of von Dassow and Odell: most tendencies found by
these authors (see Fig. 6 and Table 1 of~\cite{do02}) are confirmed by our parameter space analysis. 
There are only five exceptions, where our analysis showed no tendency (compare columns 3 and 4 
of Table~\ref{tb-par-space}), suggesting that these five parameters can, in fact, take values in a larger 
set, implying that the parameter space is larger than estimated in~\cite{do02}. 
From these exceptions, $\kappa_{\tEN\tci}$, $\kappa_{\tEN\thh}$, $\kappa_{\tCN\thh}$, 
and  $\ren$$_{\tWG}$ all belong to the group of parameters which can be freely chosen. 
The other parameter is $\kappa_{\tCI\twg}$, which depends on the 
disconnected regions, and again our analysis shows that this pair has no preferred tendency.

A more detailed examination of the conditions on $\kappa_{\tCI\twg}$ and $\kappa_{\tCN\twg}$
turns out to be very illuminating.
First, note that $\kappa_{\tCI\twg}$ and $\kappa_{\tCN\twg}$ define the five components, in the 
sense that distinct intervals for these two parameters are given in each component.
Thus, it may be expected that the distribution of these parameters varies in each region
(Table~\ref{tb-par-space2}). 
Indeed, by plotting the histograms for $\kappa_{\tCI\twg}$ and $\kappa_{\tCN\twg}$ for each region 
alone, we note that these show a marked tendency in components $\GI-\GIV$, 
for low $\kappa_{\tCI\twg}$ and high $\kappa_{\tCN\twg}$.
In contrast, the distributions of $\kappa_{\tCI\twg}$ and $\kappa_{\tCN\twg}$ for region $\GA$
alone show an opposite tendency. 
This is consistent with the fact that the volume of $\GA$ is about 95\% of $G$ and, therefore, 
it dominates the overall tendency.
Note also that, in the four components $\GI-\GIV$, it always holds that 
$\kappa_{\tCI\twg}<\kappa_{\tCN\twg}$, clearly in agreement with the tendency observed for 
our parameter sets.
In component $\GA$, the parameters $\kappa_{\tCI\twg}$ and $\kappa_{\tCN\twg}$ must satisfy
constraints that contradict those of $\GI-\GIV$, but not necessarily exactly opposite constraints
(see Table~\ref{par-regions}). Thus more freedom results for the choice of $\kappa_{\tCI\twg}$ 
and $\kappa_{\tCN\twg}$ in $\GA$. 
The tendency of $\kappa_{\tCI\twg}$ and $\kappa_{\tCN\twg}$ in $\GI-\GIV$ is, however, the opposite 
of that observed by von Dassow and Odell, a fact that can be explained once again by the 
``second missing link''. 
Indeed, since all feasible parameter sets in~\cite{dmmo00,do02} were found only after adding the 
autocatalytic {\it wingless} activation link, it can be inferred that those parameters belong to 
region $\GA$. We conclude that the parameter space is larger than estimated by von Dassow and Odell. 
\begin{table}[ht]
\caption{Comparison between the constraints identified by von Dassow and Odell~\cite{do02},
and the exact constraints given by the five regions defined above.   
Total number of parameters generated in $G$: 70026.}
\label{tb-par-space}
\begin{center}
\begin{tabular}{llll}
\hline
  Parameter    & Description & Tendency             & Tendency \\
               &             & (\cite{do02}, Table 1) & (within $G$) \\ 
\hline
$\kappa_{\tWG\ten}$    &  WG activation of en               & Moderate & Moderate \\ 
$\kappa_{\tCN\ten}$    &  CN repression of en               & Strong   & Strong \\ 
$\kappa_{\tWG\twg}$    &  WG autoactivation                 & Moderate & Moderate \\ 
$\kappa_{\tCI\twg}$    &  CI activation of wg               & Weak     & --- \\ 
$\kappa_{\tCN\twg}$    &  CN repression of wg               & Strong   & Strong \\ 
$\kappa_{\tCI\tptc}$   &  CI activation of ptc              & Strong   & Strong \\ 
$\kappa_{\tCN\tptc}$   &  CN repression of ptc              & Weak     & Weak \\ 
$\kappa_{\tEN\tci}$    &  EN  repression of ci              & Moderate & --- \\ 
$\kappa_{\tPTC\tCI}$   &  PTC stimulation of CI cleavage    & Strong   & Strong \\ 
$\kappa_{\tEN\thh}$    &  EN activation of hh               & Weak     & --- \\ 
$\kappa_{\tCN\thh}$    &  CN repression of hh               & Strong   & --- \\ 
$C_{\tCI}$             &  Maximal cleavage rate of CI       & Rapid    & Rapid \\ 
$H_{\tIWG}$            &  Half-life of intracellular WG     & Short    & Short \\ 
$\ren$$_{\tWG}$        &  Rate of WG endocytosis            & Slow     & --- \\ 
$\rex$$_{\tWG}$        &  Rate of WG exocytosis             & Moderately slow  & Moderately fast \\ 
$r_{\mbox{\tiny Mxfer}\tWG}$  & Rate of WG cell-to-cell exchange  & Slow   & Slow \\ 
$\alpha_{\tWG\twg}$    &  Maximal WG autocatalytic rate     & ---    & Moderately rapid \\ 
\hline
\end{tabular} 
\end{center}
\end{table}

\begin{table}[ht]
\caption{Influence of the autocatalytic WG activation link in the parameter distribution.}
\label{tb-par-space2}
\begin{center}
\begin{tabular}{llll}
\hline
  Parameter     & Tendency               & Tendency         & Tendency \\
                & (\cite{do02}, Table 1) &  $\GA$           & $\GI,\GII,\GII,\GIV$ \\ 
                &                        & ($\WG\rightarrow\wg$)    & ($\WG\nrightarrow\wg$)\\ 
\hline
$\kappa_{\tCI\twg}$         & Weak       & Weak            & Strong  \\ 
$\kappa_{\tCN\twg}$         & Strong     & (Moderately) Strong           & Weak  \\ 
\hline
\end{tabular} 
\end{center}
\end{table}

\section{Geometry and robustness}
The volume estimates for the parameter space regions give an idea of ``how many''
parameter combinations are possible. But volume alone is often not a reliable measure 
for robustness, as illustrated in Fig.~\ref{fig-regions-shape}.
The shape or geometry of the parameter space regions also shows how far perturbations 
around each parameter will disrupt the network.
Thus, parameter regions exhibiting ``narrow'' pieces or ``sharp'' corners indicate
a lower level of robustness in the network. 
One way to explore the shape of a given multi-dimensional set is to consider a random point ($p^0$)
and follow a random walk in space ($p^k=p^{k-1}+dp^k$, $k=1,2,\ldots,$), 
where each step has the same absolute length ($\abs{dp^k}=a_0$), but a random direction.
Then record the number of steps needed for the point to exit the given set. 
Repeating this procedure for many points in the set, the probability that a point leaves the set
after $t$ steps can be computed.


The random walk could be interpreted as parameter changes due to evolution, and the probability of
exiting after $t$ steps represents the probability that the network is no longer capable of correctly 
performing its function (for instance, when a lethal mutation occurs). Studying the first exit problem is the natural thing to do in certain evolutionary models. Suppose we consider a fitness landscape on the parameter space where the functioning regions have a fixed high fitness and every other region has zero fitness. If we consider a space of alleles to be nearly continuous and model the effect of mutation as diffusion in this space, as is often done in the adaptive dynamics literature \cite{wg05}, we find that we need to compute the mutation load, namely the rate of death from exiting the high fitness region. This idea was previously used in the context of transcriptional networks \cite{sds02}.

To explore the shape of the regions $\GI$ to $\GA$, Algorithm I uses a random walk in 
the parameter space, and checks ``exit times'' as well as the ``failed parameters''.

\vspace{3mm}
\textbf{Algorithm I}
\bit
\item[] Pick a positive number $a_0$ to be the constant magnitude of the random walk step.
\item[]   Repeat points 1-4 (run $q$),  $Q$ times. 
\item[1.] Step $0$: 
   generate a point $p^0=(p_1^0,\ldots,p_m^0)'$ at random in the parameter region $G_{\gamma}$; 
\item[2.] Step $k-1/2$, $k\geq1$: 
   generate a random perturbation $dp^{k}\in[-a_0,a_0]^m$, such that 
   $\abs{dp^{k}}=a_0$;\footnote{This step corresponds to generating a random point from a 
   uniform distribution over the hypersphere in $n$ dimensions, which can be achieved by 
   the Box and Muller transformation~\cite{bm58}. Briefly, for $i=1,\ldots,n$ pick $z_i$ randomly
   from a gaussian distribution of mean zero and variance one. Then normalize to obtain
   $z=(z_1,\ldots,z_n)/\sqrt{z_1^2+\cdots+z_n^2}$.}

\item[3.] Step $k$, $k\geq1$: 
   check if $p^k=p^{k-1}+dp^{k}$ is still in $G_{\gamma}$; 
\item[4.] Check. 
The random walk exits the parameter region at time $t$ if $p^{k}\in G_{\gamma}$ for $k<t$ 
          but $p^t\notin G_{\gamma}$. \\
   Let $p_{j_1}^t,\ldots,p_{j_J}^t$ be parameters that fail to satisfy the hierarchy of conditions which 
               defines $G_{\gamma}$.\\
   Update the exit times vector: \texttt{exit($q$)=$t$}.\\
   Update the failed parameters vector: \texttt{failpar($q$)=$[j_1,\ldots,j_J]$}.\\
\eit
To interpret the numerical results obtained with Algorithm I, define the probability
that a mutation takes place in the first $t$ steps by:
\beq
  P_{\mut}(t)  = \frac{1}{Q} \mbox{ card}(I_t),\ \ \ 
  I_t = \{q\in\N: \texttt{exit($q$)}\leq t\},
\eeq 
where card($\cdot$) denotes the cardinality of a set.
Algorithm I was applied to each component of the feasible parameter space of the segment polarity network, 
with $a_0=1\times10^{-3}$ and $Q=4000$.
Two striking facts are revealed. First,  with a significant probability, fluctuations in the 
parameters will drive the system from the operating regions $\GI$, $\GII$, $\GIII$, or $\GIV$,  
to the region $\GA$  and, conversely, switching was also observed from $\GA$ to the other four 
(see the switching column in Table~\ref{tb-percentages}).
Recalling the difference between $\GA$ and the other four components, this means that, in a 
significant number cases, the network responds to perturbations by switching to an alternative 
biological pathway, rather than break down.
A second fact is that only a very small number of parameters (six out of 39, namely
$\kappa_{\tCN\ten}$, $\kappa_{\tWG\ten}$, $\kappa_{\tCI\tptc}$, $\kappa_{\tPTC\tCI}$, 
$\kappa_{\tCI\twg}$, $\kappa_{\tCN\twg}$) 
are responsible for above 90\% of network failures or mutations. 
The percentage of cases where each of these parameters failed is shown in Table~\ref{tb-percentages}.
\begin{table}[ht]
\caption{Fragile parameters.}
\label{tb-percentages}
\begin{center}
\begin{tabular}{lccccccc}
\hline
  $G_{\gamma}$     & \% switching      & \% failed   & \% failed   & \% failed  & \% failed  & \% failed   & \% failed   \\
             &   & 
 $\kappa_{\tCN\ten}$& $\kappa_{\tWG\ten}$ & $\kappa_{\tCI\tptc}$ & $\kappa_{\tPTC\tCI}$ & $\kappa_{\tCI\twg}$ & $\kappa_{\tCN\twg}$ \\
\hline
  $\GI$     &  $1.21\%$ (to $\GA$)        &  $5.4\%$ &  $6.3\%$   &  $20.0\%$ & $4.4$ & $50.8\%$ & $5.6\%$ \\
  $\GII$    &  $0.88\%$ (to $\GA$)        &  $4.7\%$ &  $5.6\%$   &  $17.2\%$ & $4.3$ & $40.0\%$ & $18.7\%$ \\
  $\GIII$   &  $1.50\%$ (to $\GA$)        &  $6.2\%$ &  $5.2\%$   &  $30.8\%$ & $4.1$ & $35.0\%$ & $11.5\%$\\
  $\GIV$    &  $1.16\%$ (to $\GA$)        &  $4.9\%$ &  $5.3\%$   &  $16.8\%$ & $5.0$ & $40.1\%$ & $18.9\%$\\
  $\GA$     &  $0.02\%$ (to $\GI$-$\GIV$) &  $8.5\%$ &  $5.7\%$   &  $21.2\%$ & $4.8$ & $31.4\%$ & $18.8\%$\\
\hline
\end{tabular} 
\end{center}
\end{table}

Calculating the distribution function $P_{\mut}$ shows that the probability of
mutation increases very rapidly for small times, in all five components (see Fig.~\ref{fig-prob-mutation}) 
-- this indicates a low robustness of the network, because it is very likely that a very small 
number of fluctuations leads out of the feasible parameter space. 
To compare the results for the five components, we computed some quantities of interest.
A possible indicator of robustness is $T_{1/2}$, defined as the
time for which there is a 50\% chance that the system has already suffered a mutation.
Low $T_{1/2}$ indicates a system which has a low robustness to perturbations.
Another indicator  is $P_{\mut}(10)$, which gives the probability that
the system has been disrupted after only 10 perturbation steps.
Similarly, the values $P_{\mut}(100)$, $P_{\mut}(1000)$, and $P_{\mut}(10000)$ are also shown for comparison.
The computed values are summarized in Table~\ref{tb-probabilities}. 
\begin{table}[ht]
\caption{Indicators of robustness.}
\label{tb-probabilities}
\begin{center}
\begin{tabular}{lccccc}
\hline
  $G_{\gamma}$  & $T_{1/2}$ &  $P_{\mut}(10)$ & $P_{\mut}(100)$ & $P_{\mut}(1000)$ & $P_{\mut}(10000)$   \\
\hline
  $\GI$  & $18$ & $0.40$ & $0.74$ & $0.94$ & $0.99$ \\
  $\GII$ & $23$ & $0.37$ & $0.72$ & $0.94$ & $0.99$  \\
  $\GIII$& $19$ & $0.40$ & $0.72$ & $0.93$ & $0.99$\\ 
  $\GIV$ & $23$ & $0.37$ & $0.73$ & $0.94$ & $0.99$ \\
  $\GA$  & $309$ & $0.30$ & $0.41$ & $0.61$ & $0.88$ \\
\hline
\end{tabular} 
\end{center}
\end{table}
Comparison of the values for  $T_{1/2}$ and $P_{\mut}(10^d)$ ($d=1,\ldots,4$) in the five regions, 
shows clearly that the components $\GI$ to $\GIV$  result in a less robust network, 
while $\GA$ exhibits a much higher level of robustness. 
Furthermore, there is a non-negligible probability ($\approx1\%$) that
the network switches from the other components to $\GA$ instead of breaking down, 
thus contributing to the robustness level of this component.
Distinguishing the levels of robustness among the four regions $\GI$ to $\GIV$ is not straightforward,
since the indicators $P_{\mut}(10^d)$ and $T_{1/2}$ are very similar.

As noted above, only six out of 39 parameters are responsible for over 90\% of failures.
Curiously, two of these parameters satisfies constraints which are in fact independent 
of the regions ($\kappa_{\tCI\tptc}$ and $\kappa_{\tPTC\tCI}$, see Table~\ref{par-common}).
So, the numerical results clearly show that the parameter space is very narrow in the 
directions defined by $\kappa_{\tCI\tptc}$ and $\kappa_{\tPTC\tCI}$.
The other critical directions are defined by $\kappa_{\tCI\twg}$ and $\kappa_{\tCN\twg}$,
and $\kappa_{\tCN\ten}$ and $\kappa_{\tWG\ten}$,
which satisfy different conditions in each of the five parameter regions (Table~\ref{par-regions}). 
Together with common parameter $\kappa_{\tCI\tptc}$, the parameters that regulate activation and
inhibition of {\em wingless} by Cubitus proteins ($\kappa_{\tCI\twg}$ and $\kappa_{\tCN\twg}$)
are the most critical.

The main conclusion from Algorithm I clearly follows the preliminary estimates of the 
relative volumes (compare Tables~\ref{tb-volume} and~\ref{tb-probabilities}, both concluding 
that $\GA$ is more robust than the other four components). 
But the geometry analysis reveals three new fundamental results: 
(i) the system increases its robustness to environment perturbations by switching to an alternative
biological pathway. The switching event may be from a ``small'' to a ``large'' region but
also, more remarkably, from a ``large'' to a ``small''; 
(ii) the lack of robustness is due not only to small sized regions, but in part to critical
parameters ($\kappa_{\tCI\tptc}$ and $\kappa_{\tPTC\tCI}$), which define directions along which the parameter space 
is globally very narrow;
(iii) the volume alone is not a reliable measure of robustness, since volume (Table~\ref{tb-volume}) 
and the indicator $T_{1/2}$ provide different robustness classifications for components $\GI$ to $\GIV$.
For instance, the volume of $\GII$ is apparently the smallest (an indicator of low robustness), 
but $T_{1/2}$ is the largest (an indicator of high robustness), suggesting that the shape of the 
region does plays an important role. 
In contrast, the numbers $P_{\mut}(10^d)$ are very similar, suggesting that robustness levels 
of $\GI$ to $\GIV$ are in fact very similar. 
However, it should be noted that neither volume nor $T_{1/2}$ provide conclusive information
on the relative levels of robustness of  $\GI$ to $\GIV$. In particular, note that $T_{1/2}$
depends on the magnitude of the random walk step - other numerical experiments were performed with
different $a_0$ values (not shown), and the comparison results are unchanged.

\section{Discussion and conclusions}
Analysis of the feasible parameter set, by estimating its volume, identifying connected components, 
and its geometric properties are valuable tools for establishing and quantifying robustness  
in regulatory networks.
The concept of robustness, in the sense that the system's regulatory functions should operate 
correctly under a variety of situations, is closely related to the parameter space and the 
effect of parameter perturbations.
In this context, our analysis suggests that the segment polarity network is vulnerable 
to perturbations in its parameters. 
Indeed, the first striking result from our analysis is that the feasible parameter space is 
composed of five disconnected components. An implication of this topological characterization
is a diminished capacity of the network to respond well to environmental perturbations.
Random fluctuations will often drive the system to a set of parameters outside any feasible 
region, and thus lead to a break down of the network or a different phenotype.
Indeed, as the results of Algorithm I show, sucessive random perturbations to the parameters will
drive the system out of the feasible parameter set, with a large probability. 
For instance, if parameters are randomly perturbed for up to 10 times, each of magnitude 
$1\times10^{-3}$ in any direction, there is a 30\% probability that the system will fail to operate 
correctly (see Table~\ref{tb-probabilities}, column $P_{\mut}(10)$).
On the other hand, it is possible that a series of fluctuations in the environment may drive the 
system to adopt an alternative biochemical pathway, and thus ``jump'' from one feasible component 
to another (with probability 1\%, see Table~\ref{tb-percentages}).

As the group of most fragile parameters suggest, the Cubitus-{\it wingless} interactions are 
at the basis of the appearance of disconnected regions of parameters. 
Dis-connectivity in the space of parameters can be traced in large part to an incompatibility 
of Cubitus repression functions in the second cell: $\CN_2$ should be present to repress 
{\it engrailed} expression, but should be absent to enhance $\CI_2$ activation of {\it wingless}.
To increase the network's robustness to environmental fluctuations, the segment polarity model should 
account for {\it engrailed} regulation by other factor than Cubitus. One possibility is to include
regulation by pair-rule gene products, such as Sloppy paired, as explored both in~\cite{ao03} 
and ~\cite{i04}.
An external factor, again possibly from the pair-rule genes, will also play a major role in 
establishing asymmetry in the {\it cubitus} levels ($U_i$). These contribute to a larger
admissible parameter space, and together with an improved {\it engrailed} regulation, will
greatly enhance robustness of the segment polarity network in maintaining its pattern. 
An extension of the current analysis including the regulation by Sloppy paired is currently in
preparation by A. Dayarian at one of our labs~\cite{adel-ICSB07}.

Both the volume estimates and the probability of failure or mutation ($P_{\mut}$) in each component
indicate that $\GA$ is the most robust parameter region, while $\GI$, $\GII$, $\GIII$ and $\GIV$
are less robust regions, all at the same level. However, volume is not a reliable indicator
of robustness by itself, and fails to predict alternative robustness mechanisms.
Additional knowledge on the network mechanisms has been gained with the geometry analysis.
A noteworthy fact is the non-negligible probability (1\%) 
that fluctuations in the parameters in regions $\GI$ to $\GIV$ result in a switch to the 
region $\GA$, and remarkably (but with lower probability 0.02\%) also from $\GA$ to the others. 
Of the five disconnected components, $\GI$, $\GII$, $\GIII$, and $\GIV$ correspond to the pathway 
where {\it wingless} is regulated by Cubitus interruptus proteins, 
while $\GA$ corresponds to the pathway where {\it wingless} is regulated by its own protein levels.
Thus it is more likely that wild type expression in the segment polarity network is achieved through 
the Wingless auto-activation pathway. 
In the absence of the auto-activation link, von Dassow et. al. failed to observe any feasible parameter 
set in their numerical experiments. However, as soon as the auto-activation pathway was added
(the second ``missing link'' in the model~\cite{dmmo00}), immediately a significant percentage
of feasible parameter sets were observed.
This is not surprising, as elucidated by our analysis: while {\it wingless} auto-activation is not 
strictly necessary to establishing the segment polarity genes pattern, it does greatly increase 
the probably that the pattern is achieved ($\GA$ has a much larger volume, by a factor at least 40, 
and also exhibits higher robustness indices).

Another fundamental conclusion from the geometry analysis is the existence of six (out of 39) 
critical parameters which are responsible for 90\% of the network failures due to parameter 
fluctuations. Moreover, the intervals for two of these parameters ($\kappa_{\tCI\tptc}$ and $\kappa_{\tPTC\tCI}$, 
Table~\ref{par-common}) are independent of parameter space components. 
The feasible parameter set is thus globally restricted by these  parameters, which define
``narrow'' directions (see Fig.~\ref{fig-regions-shape} (b) ).

Robustness of a regulatory module should not be measured simply as a function of the volume of 
its admissible parameter space. The geometry (for instance, convexity or existence of sharp points) 
and topology (connectedness) of the parameter space play fundamental roles in measuring robustness.
The analysis developed in this paper can be applied to other systems and regulatory networks, to
systematically characterize and explore the admissible space of parameters, its
topology and geometry. These provide reliable information on how the network's interactions 
contribute to its robustness or fragility, and serve as measures to classify robust regulatory modules.

\section*{Acknowledgements}
The authors wish to specially thank Adel Dayarian for his careful checking of many computations,
as well as the Matlab codes implemented for this paper. We are very grateful for his useful comments
and corrections. One of the authors (A.M.S.) thanks Pankaj Mehta for
discussions on the segment polarity network  that lead to the formulation of
the high Hill coefficient version of the model.
E.D.S.'s work was partially supported by NSF grant DMS-0614371.
A.M.S.'s work was partially supported by a NHGRI grant R01HG03470.


\begin{thebibliography}{10}
\providecommand{\url}[1]{\texttt{#1}}
\providecommand{\urlprefix}{URL }
\expandafter\ifx\csname urlstyle\endcsname\relax
  \providecommand{\doi}[1]{doi:\discretionary{}{}{}#1}\else
  \providecommand{\doi}{doi:\discretionary{}{}{}\begingroup
  \urlstyle{rm}\Url}\fi
\providecommand{\bibAnnoteFile}[1]{%
  \IfFileExists{#1}{\begin{quotation}\noindent\textsc{Key:} #1\\
  \textsc{Annotation:}\ \input{#1}\end{quotation}}{}}
\providecommand{\bibAnnote}[2]{%
  \begin{quotation}\noindent\textsc{Key:} #1\\
  \textsc{Annotation:}\ #2\end{quotation}}
\providecommand{\eprint}[2][]{\url{#2}}

\bibitem{asbl99}
Alon U, Surette MG, Barkai N, Leibler S (1999) Robustness in bacterial
  chemotaxis.
\newblock Nature 397:168--171.
\input{asbl99}

\bibitem{lsw99}
Little J, Shepley D (1999) Robustness of a gene regulatory circuit.
\newblock EMBO J 18:4299--4307.
\input{lsw99}

\bibitem{dmmo00}
von Dassow G, Meir E, Munro E, Odell G (2000) The segment polarity network is a
  robust developmental module.
\newblock Nature 406:188--192.
\input{dmmo00}

\bibitem{sds02}
Sengupta A, Djordjevic M, Shraiman B (2002) Specificity and robustness in
  transcription control network.
\newblock Proc Natl Acad Sci USA 99:2072--2077.
\input{sds02}

\bibitem{savageau71}
Savageau M (1971) Parameter sensitivity as a criterion for evaluating and
  comparing the performance of biochemical systems.
\newblock Nature 229:542--544.
\input{savageau71}

\bibitem{heinrich96}
Heinrich R, Schuster S (1996) The regulation of cellular systems.
\newblock Chapman {\&} Hall, New York.
\input{heinrich96}

\bibitem{sanson01}
Sanson B (2001) Generating patterns from fields of cells. examples from {{\it
  Drosophila}} segmentation.
\newblock EMBO Reports 21:1083--1088.
\input{sanson01}

\bibitem{i04}
Ingolia N (2004) Topology and robustness in the {Drosophila} segment polarity
  network.
\newblock PLoS Biology 2:0805--0815.
\input{i04}

\bibitem{ao03}
Albert R, Othmer HG (2003) The topology of the regulatory interactions predicts
  the expression pattern of the {{\it Drosophila}} segment polarity genes.
\newblock J Theor Biol 223:1--18.
\input{ao03}

\bibitem{cas05}
Chaves M, Albert R, Sontag E (2005) Robustness and fragility of boolean models
  for genetic regulatory networks.
\newblock J Theor Biol 235:431--449.
\input{cas05}

\bibitem{csa06}
Chaves M, Sontag E, Albert R (2006) Methods of robustness analysis for boolean
  models of gene control networks.
\newblock IEE Proc Syst Biol 153:154--167.
\input{csa06}

\bibitem{mlot06}
Ma W, Lai L, Ouyang Q, Tang C (2006) Robustness and modular design of the
  drosophila segment polarity network.
\newblock Mol Syst Biol 2:70.
\input{mlot06}

\bibitem{do02}
von Dassow G, Odell G (2002) Design and constraints of the {\em drosophila}
  segment polarity modude: robust spatial patterning emerges from intertwined
  cell state switches.
\newblock J Exp Zool (Mol Dev Evol) 294:179--215.
\input{do02}

\bibitem{sg04}
Swantek D, Gergen JP (2004) Ftz modulates runt-dependent activation and
  repression of segment -polarity gene transcription.
\newblock Development 131:2281--2290.
\input{sg04}

\bibitem{hi90}
Hidalgo A, Ingham PW (1990) Cell patterning in the {{\it Drosophila}} segment:
  spatial regulation of the segment polarity gene {{\it patched}}.
\newblock Development 110:291--301.
\input{hi90}

\bibitem{ek90}
Eaton S, Kornberg TB (1990) Repression of ci-d in posterior compartments of
  drosophila by {{\it engrailed}}.
\newblock Genes \& Dev 4:1068--1077.
\input{ek90}

\bibitem{c75}
Collins GE (1975) Quantifier elimination for real closed fields by cylindrical
  algebraic decomposition.
\newblock In: Second GI Conference on Automata Theory and Formal Languages,
  Kaiserslauten, Springer, volume~33 of \emph{Lecture Notes Comp. Sci.} pp.
  134--183.
\input{c75}

\bibitem{acm84}
Arnon DS, Collins GE, McCallum S (1984) Cylindrical algebraic decomposition
  {I}: the basic algorithm.
\newblock SIAM J Comput 13:865--877.
\input{acm84}

\bibitem{qepad}
Brown C, Hong H, et~al.
\newblock {QEPAD}.
\newblock {http://www.cs.usna.edu/{\verb1~1}qepcad/B/QEPCAD.html}.
\input{qepad}

\bibitem{ch91}
Collins GE, Hong H (1991) Partial cylindrical algebraic decomposition in
  quantifier elimination.
\newblock J Symb Comput 12:299--328.
\input{ch91}

\bibitem{wolfram}
Wolfram S (1998) The Mathematica Book, 4th ed.
\newblock Wolfram Media, Cambridge University Press.
\input{wolfram}

\bibitem{nmgj01}
Nesi\'c D, Mareels IMY, Glad ST, Jirstrand M (2001) Software for control system
  analysis and design, symbol manipulation.
\newblock In: Webster J, editor, Encyclopedia of Electrical and Electronics
  Engineering, J. Wiley.
\input{nmgj01}

\bibitem{gt04}
Ghosh R, Tomlin C (2004) Symbolic reachable set computation of piecewise affine
  hybrid automata and its application to biological modeling: Delta-notch
  protein signaling.
\newblock IEE Trans Syst Biol 1:170--183.
\input{gt04}

\bibitem{wg05}
Waxman D, Gavrilets S (2005) 20 questions on adaptive dynamics.
\newblock J Evol Biol 18:1139--1154.
\input{wg05}

\bibitem{bm58}
Box G, Muller M (1958) A note on the generation of random normal deviates.
\newblock Ann Math Stat 29:610--611.
\input{bm58}

\bibitem{adel-ICSB07}
Chaves M, Dayarian A, Sengupta A, Sontag E.
\newblock Geometry, functionality and robustness: Exploring the parameter space
  of the segment polarity network.
\newblock Poster at {\it The 8th Int. Conf. Systems Biology}, Long Beach, CA,
  October 2007.
\input{adel-ICSB07}

\end{thebibliography}

\clearpage
\appendix

\section{Notation}
\label{sec-notation}
\begin{figure}[h]
\centerline{
\scalebox{0.8}[0.8]{\includegraphics{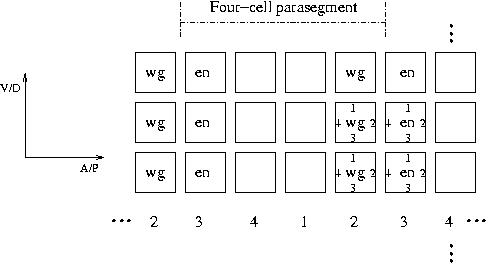} } }
\caption{Four cells in a parasegment, with periodic boundary conditions in both dimensions. 
Each cell has four membranes. 
The relative values of Wingless in each cell ($\EWG_{\underline i}$) are shown.}
\label{fig-cells}
\end{figure}

The original model can be found in~\cite{dmmo00,do02}. In order to make our work more clear,
we include the notation as well as the original equations below.
Without loss of generality (the geometry remains unchanged), each cell is assumed to 
have four faces (Fig.~\ref{fig-cells}), rather than six as in the original model~\cite{dmmo00}.
The model reproduces a parasegment of four cells and uses repetition of this group of four cells
to reproduce the embryo's anterior/posterior axis (A/P axis in Fig.~\ref{fig-cells}), and the 
circular ventral/dorsal axis (V/D axis in Fig.~\ref{fig-cells}).
Because intercellular diffusion is only considered along the A/P axis (left/right), and because 
cells repeat in the orthogonal V/D direction (up/down), it is indeed equivalent to consider symmetric
four-sided or six-sided hexagonal cells.

A saturation function, and its horizontal reflexion, are introduced:
\beq
  \phi(X,\kappa,\nu) &=& \frac{X^\nu}{\kappa^\nu+X^\nu}, \\
  \psi(X,\kappa,\nu) &=& 1-\phi(X,\kappa,\nu).
\eeq
The subscripted variables are as follows:
\beq
  X_i &=& \mbox{concentration of species $X$ on cell $i$ (when homogeneous throughout the cell ),}\\
  X_{i,j} &=& \mbox{concentration of species $X$ on cell $i$, at face $j$},\\
  \kappa_{XY} &=& \mbox{threshold for activation of species $Y$, induced by species $X$},\\
  n(i,j) &=& \mbox{index of neighbor to cell $i$, at face $j$},\\
  X_{n(i,j),j+3} &=& \mbox{concentration of species $X$ on cell face apposite to $i,j$},\\
  X_{i,\tT} &=& \sum_{j=1}^6\ X_{i,j} = \mbox{total concentration of species $X$ on cell $i$}, \\ 
  X_{\underline i} &=& \sum_{j=1}^6\ X_{n(i,j),j+3} 
            = \mbox{total concentration of species $X$ presented to cell $i$ by its neighbors}.\\
\eeq

\section{Original equations}
\label{sec-original-eqs}
From~\cite{dmmo00,do02}, the model equations are:
\beqn{eq-continuous}
  \frac{d\en_i}{dt}&=&\frac{1}{H_{\ten}}
                      \left( 
                        \phi(\EWG_{\underline i}\psi(\CN_i,\kappa_{\tCN\ten},\nu_{\tCN\ten}),
                               \kappa_{\tWG\ten},\nu_{\tWG\ten})
                        -\en_i 
                      \right) 
  \\ && \nonumber \\
  \frac{d\EN_i}{dt}&=&\frac{1}{H_{\tEN}}(\en_i-\EN_i)
  \\ && \nonumber \\
  \frac{d\wg_i}{dt}&=&\frac{1}{H_{\twg}}
  \left(
    \frac{ \alpha_{\tCI\twg}\phi(\CI_i\psi(\CN_i,\kappa_{\tCN\twg},\nu_{\tCN\twg}),
                               \kappa_{\tCI\twg},\nu_{\tCI\twg}) 
          +\alpha_{\tWG\twg}\phi(\IWG_i,\kappa_{\tWG\twg},\nu_{\tWG\twg})
         }
         {1+\alpha_{\tCI\twg}\phi(\CI_i\psi(\CN_i,\kappa_{\tCN\twg},\nu_{\tCN\twg}),
                               \kappa_{\tCI\twg},\nu_{\tCI\twg}) 
           +\alpha_{\tWG\twg}\phi(\IWG_i,\kappa_{\tWG\twg},\nu_{\tWG\twg})
         }
         -\wg_i  
  \right)
  \nonumber \\  && \label{eq-c-wg}
  \\ && \nonumber \\
  \frac{d\IWG_i}{dt}&=&\frac{1}{H_{\tWG}}
                       \left( 
                          \wg_i -\IWG_i +\ren H_{\tIWG}\EWG_{i,\tT} -H_{\tWG} \rex\IWG_i
                       \right)  \label{eq-c-IWG}
  \\ && \nonumber \\
  \frac{d\EWG_{i,j}}{dt}&=&  \frac{1}{6}\rex \IWG_i -\ren\EWG_{i,j}
                           + r_{M}(\EWG_{n(i,j),j+3}-\EWG_{i,j})\nonumber  \\
                                \label{eq-c-EWG}
                        && +r_{LM}(\EWG_{i,j-1}+\EWG_{i,j+1}-2\EWG_{i,j})
                           -\frac{\EWG_{i,j}}{H_{\tWG}}
  \\ && \nonumber \\
  \frac{d\ptc_i}{dt}&=& \frac{1}{H_{\tptc}}
                      \left( 
                        \phi(\CI_i\psi(\CN_i,\kappa_{\tCN\tptc},\nu_{\tCN\tptc}),
                               \kappa_{\tCI\tptc},\nu_{\tCI\tptc})
                        -\ptc_i 
                      \right)  \label{eq-c-ptc}
  \\ && \nonumber \\
  \frac{d\PTC_{i,j}}{dt}&=&  \frac{1}{H_{\tPTC}}
                      \left( 
                        \frac{1}{6}\ptc_i -\PTC_{i,j}
                        -\kappa_{\tPTC\tHH}H_{\tPTC}[\HH]_0\HH_{n(i,j),j+3} \PTC_{i,j}
                      \right) \nonumber \\
                      &&  +r_{LM\tPTC}(\PTC_{i,j-1}+\PTC_{i,j+1}-2\PTC_{i,j})
                         \label{eq-c-PTC}
  \\ && \nonumber \\
  \frac{d\ci_i}{dt}&=& \frac{1}{H_{\tci}}
                      \left( 
                        \phi(B_i\psi(\EN_i,\kappa_{\tEN\tci},\nu_{\tEN\tci}),
                               \kappa_{\tB\tci},\nu_{\tB\tci})
                        -\ci_i 
                      \right)  \label{eq-c-ci}
  \\ && \nonumber \\
  \frac{d\CI_i}{dt}&=& \frac{1}{H_{\tCI}}
                      \left( 
                         \ci_i -\CI_i -H_{\tCI}C_{\tCI}\CI_i
                                      \phi(\PTC_{i,\tT},\kappa_{\tPTC\tCI},\nu_{\tPTC\tCI})
                      \right)  \label{eq-c-CI}
  \\ && \nonumber \\
  \frac{d\CN_i}{dt}&=& \frac{1}{H_{\tCI}}
                      \left( 
                         H_{\tCI}C_{\tCI}\CI_i
                                      \phi(\PTC_{i,\tT},\kappa_{\tPTC\tCI},\nu_{\tPTC\tCI})
                         -\CN_i
                      \right)  \label{eq-c-CN}
  \\ && \nonumber \\
  \frac{d\hh_i}{dt}&=& \frac{1}{H_{\thh}}
                      \left( 
                        \phi(\EN_i\psi(\CN_i,\kappa_{\tCN\thh},\nu_{\tCN\thh}),
                               \kappa_{\tEN\thh},\nu_{\tEN\thh})
                        -\hh_i 
                      \right)  \label{eq-c-hh}
  \\ && \nonumber \\
  \frac{d\HH_{i,j}}{dt}&=&  \frac{1}{H_{\tHH}}
                      \left( 
                        \frac{1}{6}\hh_i -\HH_{i,j}
                        -\kappa_{\tPTC\tHH}H_{\tHH}[\PTC]_0\PTC_{n(i,j),j+3} \HH_{i,j} 
                      \right) \nonumber\\ 
                       && +r_{LM\tHH}(\HH_{i,j-1}+\HH_{i,j+1}-2\HH_{i,j})
                      \label{eq-c-HH}
\eeqn

\section{Simplified model, for large $\nu$}
\label{sec-discrete-syst}
\beqn{eq-discrete-syst}
  f_{\ten_i} &=&\frac{1}{H_{\ten}}
                      \left( 
                        \step(\EWG_{\underline i}\kappa_{\tWG\ten})\step(\kappa_{\tCN\ten}-\CN_i)
                        -\en_i 
                      \right)  \label{eq-d-en}
  \\ && \nonumber \\
  f_{\tEN_i}&=&\frac{1}{H_{\tEN}}(\en_i-\EN_i) \label{eq-d-EN}
  \\ && \nonumber \\
  f_{\twg_i}&=&\frac{1}{H_{\twg}}
  \left(
    \frac{ \alpha_{\tCI\twg}\step(\CI_i-\kappa_{\tCI\twg})\step(\kappa_{\tCN\twg}-\CN_i)
          +\alpha_{\tWG\twg}\step(\IWG_i-\kappa_{\tWG\twg})
         }
         {1+\alpha_{\tCI\twg}\step(\CI_i-\kappa_{\tCI\twg})\step(\kappa_{\tCN\twg}-\CN_i)
           +\alpha_{\tWG\twg}\step(\IWG_i-\kappa_{\tWG\twg})
         }
         -\wg_i   \label{eq-d-wg}
  \right)
  \\ && \nonumber \\
  f_{\tIWG_i}&=&\frac{1}{H_{\tWG}}
                       \left( 
                          \wg_i -\IWG_i +\ren H_{\tIWG}\EWG_{i,\tT} -H_{\tWG} \rex\IWG_i
                       \right)  \label{eq-d-IWG}
  \\ && \nonumber \\
  f_{\tEWG_{i,j}}&=&  \frac{1}{4}\rex \IWG_i -\ren\EWG_{i,j}
                           + r_{M}(\EWG_{n(i,j),j+3}-\EWG_{i,j})\nonumber  \\
                                \label{eq-d-EWG}
                        && +r_{LM}(\EWG_{i,j-1}+\EWG_{i,j+1}-2\EWG_{i,j})
                           -\frac{\EWG_{i,j}}{H_{\tWG}}
  \\ && \nonumber \\
  f_{\tptc_i}&=& \frac{1}{H_{\tptc}}
                      \left( 
                        T_i \step(\CI_i-\kappa_{\tCI\tptc})\step(\kappa_{\tCN\tptc}-\CN_i)
                        -\ptc_i 
                      \right)  \label{eq-d-ptc}
  \\ && \nonumber \\
  f_{\tPTC_{i,j}}&=&  \frac{1}{H_{\tPTC}}
                      \left( 
                        \frac{1}{4}\ptc_i -\PTC_{i,j}
                        -\kappa_{\tPTC\tHH}H_{\tPTC}[\HH]_0\HH_{n(i,j),j+3} \PTC_{i,j}
                      \right) \nonumber \\
                      &&  +r_{LM\tPTC}(\PTC_{i,j-1}+\PTC_{i,j+1}-2\PTC_{i,j})
                         \label{eq-d-PTC}
  \\ && \nonumber \\
  f_{\tci_i}&=& \frac{1}{H_{\tci}}
                      \left( 
                        \step(B_i-\kappa_{\tB\tci})\step(\kappa_{\tEN\tci}-\EN_i) -\ci_i 
                      \right)  \label{eq-d-ci}
  \\ && \nonumber \\
  f{\tCI_i}&=& \frac{1}{H_{\tCI}}
                      \left( 
                         \ci_i -\CI_i -H_{\tCI}C_{\tCI}\CI_i
                                      \step(\PTC_{i,\tT}-\kappa_{\tPTC\tCI})
                      \right)  \label{eq-d-CI}
  \\ && \nonumber \\
  f_{\tCN_i}&=& \frac{1}{H_{\tCI}}
                      \left( 
                         H_{\tCI}C_{\tCI}\CI_i
                                      \step(\PTC_{i,\tT}-\kappa_{\tPTC\tCI})
                         -\CN_i
                      \right)  \label{eq-d-CN}
  \\ && \nonumber \\
  f_{\thh_i}&=& \frac{1}{H_{\thh}}
                      \left( 
                        \step(\EN_i-\kappa_{\tEN\thh})\psi(\kappa_{\tCN\thh}-\CN_i)
                        -\hh_i 
                      \right)  \label{eq-d-hh}
  \\ && \nonumber \\
  f_{\tHH_{i,j}}&=&  \frac{1}{H_{\tHH}}
                      \left( 
                        \frac{1}{4}\hh_i -\HH_{i,j}
                        -\kappa_{\tPTC\tHH}H_{\tHH}[\PTC]_0\PTC_{n(i,j),j+3} \HH_{i,j} 
                      \right) \nonumber\\ 
                       && +r_{LM\tHH}(\HH_{i,j-1}+\HH_{i,j+1}-2\HH_{i,j})
                      \label{eq-d-HH}
\eeqn

\section{Steady state pattern}
\label{sec-steady-state}
Solving equations\rf{eq-d-en}-(\ref{eq-d-HH}) at steady state ($f=0$), 
and simplifying where possible, yields the algebraic expressions:
\beqn{eq-discrete}
   \en_i &=& \step(\EWG_{\underline i}-\kappa_{\tWG\ten})\,\step(\kappa_{\tCN\ten}-\CN_i) 
     \label{eq-discrete-en}
    \\&&  \nonumber \\
   \EN_i &=& \en_i  
    \\&&  \nonumber \\
   \wg_i &=& \frac{\alpha_{\tCI\twg}\step(\CI_i-\kappa_{\tCI\twg})\step(\kappa_{\tCN\twg}-\CN_i) 
                      +\alpha_{\tWG\twg}\step(\IWG_i-\kappa_{\tWG\twg}) }
                  {1+\alpha_{\tCI\twg}\step(\CI_i-\kappa_{\tCI\twg})\step(\kappa_{\tCN\twg}-\CN_i)
                      +\alpha_{\tWG\twg}\step(\IWG_i-\kappa_{\tWG\twg}) }  
     \label{eq-discrete-wg}
    \\&&  \nonumber \\
   \IWG_i &=& \frac{H_{\tIWG}\ren}{1+H_{\tIWG} \rex}\,\EWG_{i,T} 
          +\frac{1}{1+H_{\tIWG}\rex}\,\wg_i  
          \label{eq-discrete-IWG} 
    \\&&  \nonumber \\
   M\ \EWG &=& -\frac{1}{4}\frac{\rex}{1+H_{\tIWG}\rex}\ \widetilde{\wg} 
    \label{eq-discrete-EWG}
    \\&&  \nonumber \\
   \ptc_i &=& T_i\,\step(\CI_i-\kappa_{\tCI\tptc})\,\step(\kappa_{\tCN\tptc}-\CN_i)
     \label{eq-discrete-ptc}
    \\&&  \nonumber \\
   \PTC_{i,j}&=& \frac{1}{4}\ptc_i 
                       -\kappa_{\tPTC\tHH}H_{\tPTC}[\HH]_0\HH_{n(i,j),j+3} \PTC_{i,j}
                        \nonumber \\
                   &&    +r_{LM\tPTC}H_{\tPTC}(\PTC_{i,j-1}+\PTC_{i,j+1}-2\PTC_{i,j}) 
                    \label{eq-discrete-PTC}
    \\&&  \nonumber \\
   \ci_i &=& U_i\,\step(\kappa_{\tEN\tci}-\EN_i)  \label{eq-discrete-ci}
    \\&&  \nonumber \\
   \CI_i &=& U_i\,\frac{\step(\kappa_{\tEN\tci}-\EN_i)}
                      {1+H_{\tCI}C_{\tCI}\,\step(\PTC_{i,\tT}-\kappa_{\tPTC\tCI})}
          \label{eq-discrete-CI} 
    \\&&  \nonumber \\
   \CN_i &=& U_i\,\frac{H_{\tCI}C_{\tCI}}{1+H_{\tCI}C_{\tCI}}\,
                        \step(\kappa_{\tEN\tci}-\EN_i)\,\step(\PTC_{i,\tT}-\kappa_{\tPTC\tCI})
          \label{eq-discrete-CN} 
    \\&&  \nonumber \\
   \hh_i &=& \step(\EN_i-\kappa_{\tEN\thh})\,\step(\kappa_{\tCN\thh}-\CN_i)
     \label{eq-discrete-hh}
    \\&&  \nonumber \\
   \HH_{i,j}&=& \frac{1}{4}\hh_i 
                        -\kappa_{\tPTC\tHH}H_{\tHH}[\PTC]_0\PTC_{n(i,j),j+3} \HH_{i,j} 
                        \nonumber \\
                    &&   +r_{LM\tHH}H_{\tHH}(\HH_{i,j-1}+\HH_{i,j+1}-2\HH_{i,j})
                      \label{eq-discrete-HH}
\eeqn
$\EWG$  is a vector in $\R^{16}$ with components:
\beq
  \EWG=( && \EWG_{1,1},\EWG_{1,2},\EWG_{1,3},\EWG_{1,4},
            \EWG_{2,1},\EWG_{2,2},\EWG_{2,3},\EWG_{2,4}, \\
         && \EWG_{3,1},\EWG_{3,2},\EWG_{3,3},\EWG_{3,4},
            \EWG_{4,1},\EWG_{4,2},\EWG_{4,3},\EWG_{4,4})'
\eeq
$\widetilde{\wg}$ is also a vector in $\R^{16}$, given by the following Kronecker tensor product 
\beq
   \widetilde{\wg} &=& (\wg_1,\wg_2,\wg_3,\wg_4)'\times_{kron}(1,1,1,1)' \\
                   &=&(\wg_1,\wg_1,\wg_1,\wg_1,
                    \wg_2,\wg_2,\wg_2,\wg_2,
                    \wg_3,\wg_3,\wg_3,\wg_3,
                    \wg_4,\wg_4,\wg_4,\wg_4)'.
\eeq
Putting together the 16 equations\rf{eq-d-EWG}, 
and substituting $\IWG_i$ by its steady state expression\rf{eq-d-IWG}, 
it is not difficult to see that the matrix 
$M\in\R^{16}\times\R^{16}$ is composed of various $4\times4$ blocks, as follows:
\beqn{eq-M}
  M=\begin{pmatrix}
     E &  F_{24} & 0 & F_{42} \\
     F_{42} & E & F_{24} & 0 \\
     0 & F_{42} & E & F_{24} \\
     F_{24} & 0 & F_{42} & E 
  \end{pmatrix} 
\eeqn
where 
\beq
   E=\begin{pmatrix}
       -d & r_{LM} & r_{M} &  r_{LM} \\
       r_{LM} & -d & r_{LM} & 0\\
       r_{M} & r_{LM} & -d &  r_{LM} \\
       r_{LM} & 0 & r_{LM} & -d \\
     \end{pmatrix}\ + \
  h\ \begin{pmatrix}
       1 & 1 & 1 & 1 \\
       1 & 1 & 1 & 1 \\
       1 & 1 & 1 & 1 \\
       1 & 1 & 1 & 1 \\
     \end{pmatrix}
\eeq
with
\beq
&&   d=H_{\tIWG}^{-1}+\ren+r_{M}+2r_{LM}, \\
&&   h=\frac{1}{4}\frac{H_{\tIWG}\,\rex}{1+H_{\tIWG}\rex}\ren
\eeq
\beq
   F_{24}=\begin{pmatrix}
            0 & 0 & 0 & 0 \\ 
            0 & 0 & 0 & r_M \\ 
            0 & 0 & 0 & 0 \\ 
            0 & 0 & 0 & 0 \\ 
     \end{pmatrix}, \ \ \ 
   F_{42}=F_{24}'=
     \begin{pmatrix}
            0 & 0 & 0 & 0 \\ 
            0 & 0 & 0 & 0 \\ 
            0 & 0 & 0 & 0 \\ 
            0 & r_M & 0 & 0 \\ 
     \end{pmatrix}.
\eeq
Note that the steady state equations for $\EN$, $\IWG$, $\EWG$ and $\PTC$ are algebraic,
and in fact exact solutions can be computed from the steady state values of $\wg$ and $\ptc$.
These are discussed in more detail in the Appendices~\ref{sec-WG} and ~\ref{sec-PTC-HH}.

\paragraph{Remark:}
The parameters are as in~\cite{dmmo00}, except $T_i$ and $U_i$, which represent
the maximal values of $\ptc$ and $\ci$ (respectively), in each cell. 
These take values in the interval $[0,1]$ and generalize the possible ON values of 
$\ptc$ and $\ci$.

Note that, in the simplification from\rf{eq-c-ptc} to\rf{eq-d-ptc}, we have generalized
the equation and added distinct maximal levels of expression in each cell, given by $T_i$ 
($i=1,\ldots,4$). This allows a more accurate representation of the experimental,
which shows that {\em patched} is strongly expressed in every second and fourth cells,
weakly expressed in every first cell, and not expressed in every third cell 
(see~\cite{dmmo00} for more discussion). Thus we will consider the case: $T_1<T_2=T_4$. 

A similar generalization was made to deal with the activation of {\em cubitus interruptus}.
In von Dassow\ea~model, this is due to some external parameters $B_i$ (not governed by
a dynamical equation), with a corresponding activity threshold $\kappa_{\tB\tci}$. 
However, for more generality, and to allow  distinct maximal levels of expression in each cell,
we have replaced each of the terms $\step(B_i-\kappa_{\tB\tci})$ in\rf{eq-d-ci} by a parameter
$U_i$, $i=1,\ldots,4$\rf{eq-discrete-ci}.
Furthermore, in characterizing the set of feasible parameters, it will become clear that
allowing distinct $U_i$ enlarges the space of possible parameters, by introducing the four 
regions $\GI$ to $\GIV$.

\section{Analytically solving Wingless levels}
\label{sec-WG}
The steady states of Wingless proteins\rf{eq-discrete-EWG} and\rf{eq-discrete-IWG} 
are given directly by algebraic equations,
depending only on wingless mRNA ($\wg_2$) and diffusion parameters for
intracellular (membrane-to-membrane) and  intercellular communication.
Consider equation\rf{eq-discrete-EWG}: it is easy to see that $M$ is in fact always invertible 
(if all parameters are positive). 
First note that the matrix is diagonally dominant, by adding up the entries in any column:
\beq
   -\left(H_{\tIWG}^{-1}+\ren+r_{M}+2r_{LM}\right) 
   +2r_{LM}+r_M+4h = -H_{\tIWG}^{-1} -\ren\,\frac{1}{1+H_{\tIWG}\rex}
\eeq
which is always a negative quantity. By Ger\v sgorin's Theorem, all eigenvalues of $M$ are 
contained in the disk centered at $-d+h$ with radius $2r_{LM}+r_M+3h$, therefore all have 
negative real parts. 
Thus, the matrix $M$ is symmetric and negative definite, and since the 
right-hand-side vector is also non-positive, all solutions are real and positive,
{\it whatever the choice of parameters}.
As a fact, note that the vector $\vec1=(1,1,\ldots,1)'\in\R^{16}$ is an eigenvector of $M$, 
corresponding to the eigenvalue $\lambda_1=-H_{\tIWG}^{-1} -\ren\,\frac{1}{H_{\tIWG}+\rex}$.

\paragraph{Proof of Theorem~\ref{th-EWG-symmetry}}
Assume that $\wg=(0,w,0,0)$, for any positive constant $w$. 
From the symmetry of the matrix equation\rf{eq-discrete-EWG}, several facts
can be deduced, which lead to the main result\rf{eq-EWG-symmetry}. 

\begin{fact}\label{fc-i13}
For all $i=1,2,3,4$ it holds that
\beq
  \EWG_{i,1}=\EWG_{i,3}. 
\eeq
\end{fact}
\proof
This is easy to see from the respective equations:
\beq
   (-d+h)\EWG_{i,1} + (r_{M}+h)\EWG_{i,2} +(r_M+h)\EWG_{i,3} + (r_{M}+h)\EWG_{i,4} 
                                                  = -\frac{h}{\ren H_{\tIWG}} \wg_i\\
   (r_M+h)\EWG_{i,1} + (r_{M}+h)\EWG_{i,2} +(-d+h)\EWG_{i,3} + (r_{M}+h)\EWG_{i,4} 
                                                  = -\frac{h}{\ren H_{\tIWG}} \wg_i
\eeq
which can be rearranged to
\beqn{eq-i13}
   -(d+r_M)\EWG_{i,1} + (r_{M}+h)(\EWG_{i,2}+\EWG_{i,4}) +(r_M+h)(\EWG_{i,3}+\EWG_{i,1})
                                                  = -\frac{h}{\ren H_{\tIWG}} \wg_i
   \nonumber\\
   -(d+r_M)\EWG_{i,3} + (r_{M}+h)(\EWG_{i,2}+\EWG_{i,4}) +(r_M+h)(\EWG_{i,3}+\EWG_{i,1})
                                                  = -\frac{h}{\ren H_{\tIWG}} \wg_i\ .
\eeqn
Subtracting these two equations yields the desired result.
\qed

\begin{fact}\label{fc-indexes}
It holds that
\beq
   \EWG_{2,2}=\EWG_{2,4},\ \ \EWG_{4,2}=\EWG_{4,4}, \ \
   \EWG_{1,2}=\EWG_{3,4},\ \ \EWG_{1,4}=\EWG_{3,2}.
\eeq
\end{fact}
\proof
Exchanging the indexes:
\beq
   2,2 \leftrightarrow 2,4\ \ \     4,2 \leftrightarrow 4,4\ \ \ 
   1,2 \leftrightarrow 3,4\ \ \     1,4 \leftrightarrow 3,2
\eeq
it is easy to see that the system remains unchanged (see also Fig.~\ref{fig-cells}).
\qed

\vspace{2mm}
The equality part in\rf{eq-EWG-symmetry} is now clear:
\begin{fact}\label{fc-1=3}
$
   \EWG_{\underline1}=\EWG_{\underline3}.
$
\end{fact}
\proof
We first show that $\EWG_{1,1}=\EWG_{3,3}$.
Writing equation\rf{eq-i13} for $i=1$ and $i=3$:
\beq
-(d+r_M)\EWG_{1,1} + (r_{M}+h)(\EWG_{1,2}+\EWG_{1,4}) +(r_M+h)(\EWG_{1,3}+\EWG_{1,1})
                                                  = 0\\
-(d+r_M)\EWG_{3,3} + (r_{M}+h)(\EWG_{3,2}+\EWG_{3,4}) +(r_M+h)(\EWG_{3,3}+\EWG_{3,1})
                                                  = 0
\eeq
Using Fact~\ref{fc-i13} one has $\EWG_{1,1}=\EWG_{1,3}$ and $\EWG_{3,1}=\EWG_{3,3}$, 
and then using Fact~\ref{fc-indexes} obtains:
\beq
-(d+r_M-2r_M-2h)\EWG_{1,1} + (r_{M}+h)(\EWG_{3,4}+\EWG_{3,2}) &=& 0\\
-(d+r_M-2r_M-2h)\EWG_{3,3} + (r_{M}+h)(\EWG_{3,2}+\EWG_{3,4}) &=& 0.
\eeq
Subtracting these two equations shows that $\EWG_{1,1}=\EWG_{3,3}$.
Now recalling the notation for $X_{\underline i}$ from Appendix~\ref{sec-notation}
\beq
  \EWG_{\underline1}&=&\EWG_{1,1}+\EWG_{2,4}+\EWG_{1,3}+\EWG_{4,2} \\
  \EWG_{\underline3}&=&\EWG_{3,1}+\EWG_{4,4}+\EWG_{3,3}+\EWG_{2,2}. 
\eeq
Using $\EWG_{1,1}=\EWG_{3,3}$, Fact~\ref{fc-i13} and Fact~\ref{fc-indexes} obtains:
\beq
  \EWG_{\underline1}=\EWG_{3,1}+\EWG_{2,2}+\EWG_{3,3}+\EWG_{4,4} = 
  \EWG_{\underline3}.
\eeq
as we wanted to prove.
\qed

\vspace{2mm}
To show the other inequalities, note first that the  16 variables $\EWG_{i,j}$ 
are thus reduced to only seven:
\beq
  E_{1,1} && =\EWG_{1,1}=\EWG_{1,3}=\EWG_{3,1}=\EWG_{3,3} \\
  E_{1,2} && =\EWG_{1,2}=\EWG_{3,4} \\
  E_{1,4} && =\EWG_{1,4}=\EWG_{3,2} \\
  E_{2,1} && =\EWG_{2,1}=\EWG_{2,3} \\
  E_{2,2} && =\EWG_{2,2}=\EWG_{2,4} \\
  E_{4,1} && =\EWG_{4,1}=\EWG_{4,3} \\
  E_{4,2} && =\EWG_{4,2}=\EWG_{4,4} 
\eeq
and satisfy the equations:
\beqn{eq-E11}
  -(d-r_{M}-2h)E_{1,1}+(r_{LM}+h)E_{1,2}+(r_{LM}+h)E_{1,4}&=&0 \\
  2(r_{LM}+h)E_{1,1} -(d-h)E_{1,2} +h E_{1,4} + r_{M}E_{2,2}&=&0 \label{eq-E12}\\
  2(r_{LM}+h)E_{1,1} +h E_{1,2} -(d-h)E_{1,4} + r_{M}E_{4,2}&=&0 \label{eq-E14}\\
  -(d-r_{M}-2h)E_{2,1}+2(r_{LM}+h)E_{2,2}&=&-\frac{h}{\ren H_{\tIWG}} \wg_2 \label{eq-E21}\\
  2(r_{LM}+h)E_{2,1} -(d-2h)E_{2,2} +r_{M} E_{1,2}&=&-\frac{h}{\ren H_{\tIWG}} \wg_2  \label{eq-E22}\\
  -(d-r_{M}-2h)E_{4,1}+2(r_{LM}+h)E_{4,2}&=&0 \label{eq-E41}\\
  2(r_{LM}+h)E_{4,1} -(d-2h)E_{4,2} +r_{M} E_{1,4}&=&0\ . \label{eq-E42}
\eeqn
To simplify notation, set:
\beq
  A=d-r_{M}-2h,\ \ B=2(r_{LM}+h),\ \ 
  \bar w = \frac{h}{\ren H_{\tIWG}} \wg_2,
\eeq
and note that $A>B>0$. 

\begin{fact}\label{fc-ineq1}
The following hold:
\bit
\item[(a)] $E_{4,1}<E_{4,2}<E_{1,4}<E_{1,2}<E_{2,2}$;
\item[(b)] $E_{4,1}<E_{1,1}<E_{1,2}$;
\item[(c)] $E_{1,2}+E_{1,4}<E_{2,2}+E_{4,2}$
\eit
\end{fact}

\proof
To prove part (a), from eqs.\rf{eq-E41},\rf{eq-E42} it holds that
\beq
  E_{4,1}=\frac{B}{A}E_{4,2};\ \ \   
  E_{4,2}=\frac{r_{M}A}{A^2-B^2+r_{M}A}E_{1,4}
\eeq
Because $A>B>0$, it is clear that $E_{4,1}<E_{4,2}<E_{1,4}$.
From eqs.\rf{eq-E21},\rf{eq-E22} it holds that
\beq
  E_{2,1}=\frac{B}{A}E_{2,2} +\frac{1}{A}\bar w;\ \ \   
  E_{2,2}=\frac{r_{M}A}{A^2-B^2+r_{M}A}E_{1,2} +\frac{A+B}{A^2-B^2+r_{M}A}\bar w
\eeq
Then eqs.\rf{eq-E12},\rf{eq-E14} can be written in the form
\beq
  \left(d-r_M\frac{r_{M}A}{A^2-B^2+r_{M}A}\right)E_{1,2}&=&  
  B E_{1,1} +h(E_{1,2}+E_{1,4}) +r_M\frac{A+B}{A^2-B^2+r_{M}A}\bar w \\
  \left(d-r_M\frac{r_{M}A}{A^2-B^2+r_{M}A}\right)E_{1,4}&=&  
  B E_{1,1} +h(E_{1,2}+E_{1,4}) 
\eeq
which implies that $E_{1,4}<E_{1,2}$ (it is easy to see that the factor 
multiplying both $E_{1,2}$ and $E_{1,4}$ is positive, since $d>r_M$).

We still need to prove the last inequality in (a), but we can now prove (b). 
From eq.\rf{eq-E11}
\beq 
  E_{1,1}=\frac{1}{2}\frac{B}{A}(E_{1,2}+E_{1,4})<E_{1,2}
\eeq
using (a) and because $B<A$. This proves the second inequality in (b).
To prove (c), substitute this $E_{1,1}$ expression into the sum of 
eqs.\rf{eq-E12},\rf{eq-E14}:
\beq
  E_{2,2}+E_{4,2}=\frac{A^2-B^2+r_{M}A}{r_{M}A}(E_{1,2}+E_{1,4})>E_{1,2}+E_{1,4}.
\eeq
The last part of (a) now follows from (c) together with $E_{4,2}<E{1,4}$, 
which implies $E_{1,2}<E_{2,2}$. 

Finally, the first part of (b) is easy to see from:
\beq
  E_{1,1}-E_{4,1} = \frac{1}{2}\frac{B}{A}(E_{1,2}+E_{1,4}) -\frac{B}{A}E_{4,2}
                  >\frac{1}{2}\frac{B}{A}(E_{1,2}+E_{1,4}-E_{1,4})>0.
\eeq
\qed

\vspace{2mm}
To prove the first inequality of Theorem~\ref{th-EWG-symmetry} is now straighforward.
\begin{fact}\label{fc-4<1}
    $\EWG_{\underline4}<\EWG_{\underline1}$
\end{fact}
\proof
Recall the notation for $\EWG_{\underline i}$ and use Fact~\ref{fc-ineq1}
\beq
  \EWG_{\underline1}-\EWG_{\underline4}&=&
  2E_{1,1}+E_{2,2}+E_{4,2} -2E_{4,1}-2E_{1,4} \\
 &=&2(E_{1,1}-E_{4,1})+ (E_{2,2}+E_{4,2}-E_{1,2}-E_{1,4})+(E_{1,2}-E_{1,4})>0.
\eeq 
\qed

\vspace{2mm}
The next result finishes the proof of Theorem~\ref{th-EWG-symmetry}.
\begin{fact}\label{fc-1<3}
    $\EWG_{\underline3}<\EWG_{\underline2}$
\end{fact}

\proof
Consider:
\beq
  \EWG_{\underline2}-\EWG_{\underline1}&=&
  2E_{2,1}+2E_{1,2} -2E_{1,1}-E_{2,2}-E_{4,2} \\
 &=&2(E_{1,2}-E_{1,1})+ (E_{2,1}-E_{2,2}) +(E_{2,1}-E_{4,2}),
\eeq 
which is positive if $E_{2,1}>E_{2,2}$. We will show that this is indeed the case. 
It will be useful to see that 
\beq
   E_{1,2} = r_M\frac{A+B}{d(A^2-B^2+r_MA)-r_M^2A}\bar w
            +\frac{1}{2}r_M\frac{A^2-B^2+r_MA}{d(A^2-B^2+r_MA)-r_M^2A}
             \frac{-A^2+B^2+(d-r_M)A}{(A-B)(A^2-B^2+2r_MA)}\bar w.
\eeq
Now consider
\beq
  E_{2,1}-E_{2,2}=-\frac{A-B}{A}\frac{r_MA}{A^2-B^2+r_MA}E_{1,2}
                  -\frac{A-B}{A}\frac{A+B}{A^2-B^2+r_MA}\bar w +\frac{1}{A}\bar w. 
\eeq
The last two terms can be combined into
\beq
    \frac{r_M}{A^2-B^2+r_MA}\bar w,
\eeq
and the two terms due to $E_{1,2}$ can be simplified to:
\beq
  -r_M\frac{1}{A^2-B^2+r_MA}\,\frac{(A-B)(A+B)}{\frac{d}{r_M}(A^2-B^2+r_MA)-r_MA}\bar w
\eeq
and
\beq
  -\frac{r_M}{2}\frac{1}{A^2-B^2+2r_MA}\,\frac{-A^2+B^2+(d-r_M)A}{\frac{d}{r_M}(A^2-B^2+r_MA)-r_MA}\bar w.
\eeq
Factoring out $r_M\bar w/(\frac{d}{r_M}(A^2-B^2+r_MA)-r_MA)$, one obtains
\beq
 && \frac{1}{r_M\bar w}\frac{d}{r_M}(A^2-B^2+r_MA)-r_MA)(E_{2,1}-E_{2,2})=\\
 && \frac{1}{A^2-B^2+r_MA}\left(\frac{d}{r_M}(A^2-B^2+r_MA)-r_MA - (A^2-B^2) \right)\
  -\frac{1}{2}\frac{-A^2+B^2+(d-r_M)A}{A^2-B^2+2r_MA}
\eeq
which can be further simplified to
\beq
  \frac{\left(\frac{d}{r_M}-1\right)(A^2-B^2)}{A^2-B^2+r_MA}
  +\frac{1}{2}\frac{A^2-B^2}{A^2-B^2+2r_MA}
  +\frac{(d-r_M)A}{A^2-B^2+r_MA}-\frac{1}{2}\frac{(d-r_M)A}{A^2-B^2+2r_MA} >0
\eeq
because the first two terms are clearly positive, and the last two terms add up to a positive number.  
This shows that $E_{2,1}>E_{2,2}$, as we wanted to prove.
\qed

\section{Analytically solving $\PTC$ and $\HH$ levels}
\label{sec-PTC-HH}
In this section, we prove uniqueness of solutions for $\PTC$ and $\HH$ in the conditions 
of Theorem~\ref{th-unique-ss}. 
The steady state levels of Patched and Hedgehog proteins are given by a system of nonlinear 
equations\rf{eq-discrete-PTC} and\rf{eq-discrete-HH}. 
These equations can be solved explicitly and uniquely in the case $\ptc_2=\ptc_4=T_2$, 
which is true is the steady state output is in  $\Y^{\tWT}$.
To simplify notation, we use
\beq
  r_P=r_{LM\tPTC},\ \ r_H=r_{LM\tHH},\ \ 
  \kappa_{H}=\kappa_{\tPTC\tHH}[\HH]_0,\ \  
  \kappa_{P}=\kappa_{\tPTC\tHH}[\PTC]_0,
\eeq
and define
\beq
   d_P=\frac{1}{H_{\tPTC}} +2r_P,\ \ \
   d_H=\frac{1}{H_{\tHH}} +2r_H.
\eeq
We introduce further notation:
\beq
   \beta_P=\frac{2r_P^2d_P}{d_P^2-2r_P^2},\ \ \ \
   \gamma_P=\frac{1}{4H_{\tPTC}}+\frac{1}{4H_{\tPTC}}\frac{2r_P(r_P+d_P)}{d_P^2-2r_P^2}
           =\frac{1}{4H_{\tPTC}}\frac{d_P(2r_P+d_P)}{d_P^2-2r_P^2}.
\eeq

\begin{lemma}\label{lm-PTC-HH}
Let $x\in\WTset$ be such that $\out(x)\in\Y^{\tWT}$. 
Then, the solution for $\HH$ is:
\beq
   \HH_{i,1}=\HH_{i,2}=\HH_{i,3}=\HH_{i,4}&=&0,\ \ \ i=1,2,4,\\
   \HH_{3,2}=\HH_{3,4}&=&\texttt{Root}_+ \, ,\\
   \HH_{3,1}=\HH_{3,3}&=&\frac{1}{d_H}(\frac{1}{4}\frac{\hh_3}{H_{\tHH}} +r_H\HH_{3,2}+r_H\HH_{3,4}),
\eeq
where \texttt{Root}$_+$ is the positive root of the quadratic equation:
\beq
   k_H(d_H^2-4r_H^2)X^2 +\left( 
             (d_P-\beta_P)(d_H^2-4r_H^2) -k_H(d_H+2r_H)
                                     \frac{\hh_3}{4H_{\tHH}}+d_Hk_P\gamma_P\ptc_2 
                                      \right)X   && \\
    -(d_P-\beta_P)(d_H+2r_H)\frac{\hh_3}{4H_{\tHH}} = 0.  &&
\eeq
And the solution for $\PTC$ is:
\beq
   \PTC_{3,1}=\PTC_{3,2}=\PTC_{3,3}=\PTC_{3,4}&=&0,\\
   \PTC_{2,2}=\PTC_{4,4}&=&\frac{\gamma_P T_2}{d_P-\beta_P+k_H\HH_{3,4}},\\
   \PTC_{2,1}=\PTC_{2,3}=\PTC_{4,1}=\PTC_{4,3}
               &=& \frac{1}{d_P^2-2r_P^2}\left(
                        r_Pd_P\PTC_{2,2} 
                       +\frac{1}{4H_{\tPTC}}(d_P+r_P) T_2 \right), \\
  \PTC_{2,4}=\PTC_{4,2} &=&\frac{1}{d_P}\,(\frac{1}{4} \frac{T_{2}}{H_{\tPTC}}+2r_P\PTC_{2,1}). \\
\eeq
\end{lemma}

{\it Proof.}
Let $x\in\WTset$ and $\out(x)$ be a vector in $\Y^{\tWT}$, defined by\rf{eq-output}.
Because {\em hedgehog} is not expressed in cells 1, 2 and 4, note that for $i=1,2,4$
\beq
   \HH_{i,\tT}=\sum_{j=1}^4\ \HH_{i,j} &=& 
   \hh_i- \kappa_P(\cdots) +r_H\sum_{j=1}^4\ (\HH_{i,j-1}+\HH_{i,j-1}-2\HH_{i,j})\\
    &=& - \kappa_P(\cdots) 
\eeq
since $\hh=(0,0,1,0)$, and the sum that multiplies $r_H$ cancels out.
The terms in $\kappa_P(\cdots)$ are all nonnegative, and therefore they can only be zero.
We conclude that:
\beq
   \HH_{i,1}=\HH_{i,2}=\HH_{i,3}=\HH_{i,4}=0,\ \ \ i=1,2,4.
\eeq
A similar argument shows that $\ptc_3=0$ implies:
\beq
   \PTC_{3,1}=\PTC_{3,2}=\PTC_{3,3}=\PTC_{3,4}=0.
\eeq
Therefore, the only nonlinear terms appear in the equations for $\PTC_{2,2}$ and $\PTC_{4,4}$:
\beq
  d_P\PTC_{2,2} -r_P\PTC_{2,1} -r_P\PTC_{2,3} 
           +\kappa_{H}\PTC_{2,2}\HH_{3,4} &=& \frac{1}{4H_{\tPTC}}\ptc_2 \\
  d_P\PTC_{4,4} -r_P\PTC_{4,1} -r_P\PTC_{4,3} 
           +\kappa_{H}\PTC_{4,4}\HH_{3,2} &=& \frac{1}{4H_{\tPTC}}\ptc_4\ . \\
\eeq
Moreover, symmetry of the system shows that $\PTC_{2,1}=\PTC_{2,3}$ and  $\PTC_{4,1}=\PTC_{4,3}$,
because each pair satisfies exactly the same equation:
\beqn{eq-PTCt}
  d_P\PTC_{2,1} -r_P\PTC_{2,2} -r_P\PTC_{2,4}&=& \frac{1}{4H_{\tPTC}}\ptc_2\\
  d_P\PTC_{4,3} -r_P\PTC_{4,4} -r_P\PTC_{4,2}&=& \frac{1}{4H_{\tPTC}}\ptc_4.
\eeqn
We then have:
\beq
  \PTC_{2,4} &=&\frac{1}{d_P}\,(\frac{1}{4H_{\tPTC}}\ptc_{2}+2r_P\PTC_{2,1}) \\
  \PTC_{4,2} &=&\frac{1}{d_P}\,(\frac{1}{4H_{\tPTC}}\ptc_{4}+2r_P\PTC_{4,1}). \\
\eeq
Solving for $\PTC_{2,1}$ as a function of $\PTC_{2,2}$, 
and for $\PTC_{4,1}$ as a function of $\PTC_{4,4}$:
\beq
  \PTC_{2,1} &=& \frac{1}{d_P^2-2r_P^2}\left(
                        r_Pd_P\PTC_{2,2} 
                       +\frac{1}{4H_{\tPTC}}(d_P+r_P)\ptc_2 \right) \\
  \PTC_{4,1} &=& \frac{1}{d_P^2-2r_P^2}\left(
                        r_Pd_P\PTC_{4,4} 
                       +\frac{1}{4H_{\tPTC}}(d_P+r_P)\ptc_4 \right). \\
\eeq
Thus we get equations depending only on $\PTC_{2,2}$ and $\HH_{3,4}$, 
and on $\PTC_{4,4}$ and $\HH_{3,2}$:
\beqn{eq-PTC2244}
  d_P\PTC_{2,2} -\frac{2r_P}{d_P^2-2r_P^2}
        \left(  r_Pd_P\PTC_{2,2} +\frac{1}{4H_{\tPTC}}(d_P+r_P)\ptc_2 \right)
        +\kappa_{H}\PTC_{2,2}\HH_{3,4} \,=\, \frac{1}{4H_{\tPTC}}\ptc_2 && \\ && \nonumber \\
  d_P\PTC_{4,4} -\frac{2r_P}{d_P^2-2r_P^2}
        \left(  r_Pd_P\PTC_{4,4} +\frac{1}{4H_{\tPTC}}(d_P+r_P)\ptc_4\right)  
        +\kappa_{H}\PTC_{4,4}\HH_{3,2} \,=\, \frac{1}{4H_{\tPTC}}\ptc_4\ . && \label{eq-PTC2244b}\\
       \nonumber
\eeqn
On the other hand, since $\PTC_{3,j}=0$ for all $j$, it follows that:
\beq
  \HH_{3,1}=\HH_{3,3}=\frac{1}{d_H}
            (\frac{1}{4H_{\tHH}}\hh_3 +r_H\HH_{3,2}+r_H\HH_{3,4}),
\eeq
and substituting into the $\HH_{3,4}$ and $\HH_{3,2}$ equations:
\beqn{eq-HH3234}
  d_H\HH_{3,4}-2\frac{r_H}{d_H}
            (\frac{1}{4H_{\tHH}}\hh_3 +r_H\HH_{3,2}+r_H\HH_{3,4}) -\kappa_{P}\PTC_{2,2}\HH_{3,4} 
            = \frac{1}{4H_{\tHH}}\hh_3 && \\
  d_H\HH_{3,2}-2\frac{r_H}{d_H}
            (\frac{1}{4H_{\tHH}}\hh_3 +r_H\HH_{3,2}+r_H\HH_{3,4}) -\kappa_{P}\PTC_{4,4}\HH_{3,2} 
            = \frac{1}{4H_{\tHH}}\hh_3\ .  && \label{eq-HH3234b} \\ \nonumber
\eeqn
The last four equations may be solved for the four variables $\PTC_{2,2}$, $\PTC_{4,4}$
$\HH_{3,2}$ and $\HH_{3,4}$, and the remaining $\PTC$, $\HH$ will then follow.
Recalling the notation introduced above, one can write
\beqn{eq-PTC}
   \PTC_{2,2}=\frac{\gamma_P\ptc_2}{d_P-\beta_P+k_H\HH_{3,4}},\ \ \ 
   \PTC_{4,4}=\frac{\gamma_P\ptc_4}{d_P-\beta_P+k_H\HH_{3,2}}.
\eeqn
This leads to
\beq
  d_H\HH_{3,4}-2\frac{r_H}{d_H}(\frac{1}{4H_{\tHH}}\hh_3 +r_H\HH_{3,2}+r_H\HH_{3,4}) 
            -\kappa_{P}\gamma_P\ptc_2\frac{\HH_{3,4}}{d_P-\beta_P+k_H\HH_{3,4}} 
            = \frac{1}{4H_{\tHH}}\hh_3  \\
  d_H\HH_{3,2}-2\frac{r_H}{d_H}(\frac{1}{4H_{\tHH}}\hh_3 +r_H\HH_{3,2}+r_H\HH_{3,4}) 
           -\kappa_{P}\gamma_P\ptc_4\frac{\HH_{3,2}}{d_P-\beta_P+k_H\HH_{3,2}}
            = \frac{1}{4H_{\tHH}}\hh_3\ .  \\   
\eeq
From the symmetry of these equations, it is easy to see that 
\beq
   \ptc_2=\ptc_4 \ \ \ \Rightarrow\ \ \ \HH_{3,4}=\HH_{3,2}.
\eeq
and thus have the following equation for
$\HH_{3,4}=\HH_{3,2}=X$ (after some simple algebra steps):
\beqn{eq-HHpoly}
   k_H(d_H^2-4r_H^2)X^2 +\left( 
             (d_P-\beta_P)(d_H^2-4r_H^2) -k_H(d_H+2r_H)
                                     \frac{\hh_3}{4H_{\tHH}}+d_Hk_P\gamma_P\ptc_2 
                                      \right)X   && \nonumber \\
    -(d_P-\beta_P)(d_H+2r_H)\frac{\hh_3}{4H_{\tHH}} = 0.  &&
\eeqn 
We next show that only one of the two roots of this second order polynomial is positive and
hence the unique solution to $\HH_{3,2}$, $\HH_{3,4}$. Let the polynomial be of the form
$c_2X^2+c_1X+c_0=0$. The term inside the square root will be $c_1^2-4c_0c_2$ where:
\beq
   -4c_0c_2=4k_H(d_H^2-4r_H^2)(d_P-\beta_P)(d_H+2r_H)\frac{\hh_3}{4H_{\tHH}}.
\eeq
The factor $d_H^2-4r_H^2$ is positive, by definition of $d_H$. The factor
\beq
  d_P-\beta_P=d_P-\frac{2r_P^2d_P}{d_P^2-2r_P^2}=d_P(d_P^2-4r_P^2)
\eeq
is also positive, again by definition of $d_P$.
This means that $c_1^2-4c_0c_2>c_1^2$, so whatever the sign of $c_1$, $-c_1-\sqrt{c_1^2-4c_0c_2}<0$,
which leaves us with:
\beq
  \HH_{3,2}=\HH_{3,4}=\frac{-c_1+\sqrt{c_1^2-4c_0c_2}}{2c_2}
\eeq
(the coefficients are as in~(\ref{eq-HHpoly})).
\qed

\subsection{Asymmetry in {\it patched} ON levels}
\label{sec-ptc2-ptc4}
The assumption $T_2=T_4$ is now relaxed, and the more general case is analyzed.
The main question is how Patched asymmetry influences the space of parameters, $G$, 
and whether the five components can become connected. 
In other words, does the more general case assumption $T_2\neq T_4$ leads to a increasing 
network robustness. It will be seen that this is actually not true. 
The presence of $\CN$ in the first cell is still necessary (because Wingless protein expression is 
not affected by $\ptc$ levels), but expression of $\CN$ in the second and fourth cells may now be 
different.
While it is now difficult to explicitly solve the nonlinear equations for $\PTC_i$ and $\HH_i$, it can
still be shown that $\ptc_2<\ptc_4$ implies $\PTC_2<\PTC_4$.

\begin{fact}\label{f-ptc2-ptc4}
$(\ptc_2-\ptc_4)(\PTC_2-\PTC_4)>0$.
\end{fact}

{\it Proof.}
To see this assume that $\ptc_2>\ptc_4$ (the opposite case follows a similar argument).
From the discussion above, the Hedgehog values must satisfy
\beqn{eq-HH3432}
   d_H\HH_{3,4} - a_1\ptc_2\frac{\HH_{3,4}}{a_2+a_3\HH_{3,4}}
 = d_H\HH_{3,2} - a_1\ptc_4\frac{\HH_{3,2}}{a_2+a_3\HH_{3,2}} 
\eeqn
with some positive constants $a_{1,2,3}$.
Because this is an increasing function of $\HH_{\cdot,\cdot}$, and decreasing with $\ptc_{\cdot}$,
it follows that $\HH_{3,4}>\HH_{3,2}$. Rewriting\rf{eq-HH3432}
\beq
   \HH_{3,4}(d_H - a_1\ptc_2\frac{1}{a_2+a_3\HH_{3,4}})
 = \HH_{3,2}(d_H - a_1\ptc_4\frac{1}{a_2+a_3\HH_{3,2}}) 
\eeq
and comparing with the Patched values from\rf{eq-PTC2244}, 
\beq
   \frac{\HH_{3,4}}{\HH_{3,2}}\frac{d_H-a_0\PTC_{4,4}}{d_H-a_0\PTC_{2,2}}>1
\eeq  
for an appropriate positive constant $a_0$. This last inequality shows that
$\PTC_{2,2}>\PTC_{4,4}$. Finally, retracing back to\rf{eq-PTCt}, it is not difficult to
see that
\beq
  \ptc_2>\ptc_4 \ \ \Rightarrow\ \ \PTC_{2,\tT}>\PTC_{4,\tT}.
\eeq
\qed

\paragraph{Distinct $\ptc_2$, $\ptc_4$ does not increase robustness}
On the whole, there are four possibilities to consider: (i) $\PTC_{2,4}>\kappa_{\tPTC\tCI}$;
(ii) $\PTC_{2}>\kappa_{\tPTC\tCI}>\PTC_4$; (iii) $\PTC_{4}>\kappa_{\tPTC\tCI}>\PTC_2$; and
(iv)  $\kappa_{\tPTC\tCI}>\PTC_{2,4}$. 
As already mentioned, $\PTC_1>\kappa_{\tPTC\tCI}$ in all four situations.

Situation (i) is similar to the case $T_2=T_4$ already studied, where $\CI$ and $\CN$ have the 
form\rf{eq-CI-CN}. 
In case (ii), the Cubitus proteins have the form:
\beq
   \begin{array}{ll}
     \CI_{1,2}=U_{1,2}\frac{1}{1+H_{\tCI}C_{\tCI}}, & 
        \CN_{1,2}=U_{1,2}\frac{H_{\tCI}C_{\tCI}}{1+H_{\tCI}C_{\tCI}} \\
     \CI_4=U_4, & \CN_4=0.
   \end{array}
\eeq
The conditions for $\wg$ activation by $\CI$ in the second cell require $U_2>U_4$,
so the parameters $\kappa_{\tCI\twg}$, $\kappa_{\tCN\twg}$ may take values only from 
components $\GI$ or $\GII$, or $\GA$.
In case (iii) the Cubitus repressor protein is not present in the second cell:
\beq
   \begin{array}{ll}
     \CI_{1,4}=U_{1,4}\frac{1}{1+H_{\tCI}C_{\tCI}}, &
       \CN_{1,4}=U_{1,4}\frac{H_{\tCI}C_{\tCI}}{1+H_{\tCI}C_{\tCI}}\\
     \CI_2=U_2, & \CN_2=0.
   \end{array}
\eeq
Since $\CN_2=0$, repression of {\it engrailed} on the second cell must 
now be due to insufficient Wingless activation, implying:
\beq
   \EWG_{\underline2}<\kappa_{\tWG\ten}<\EWG_{\underline3}
\eeq
which is impossible, since it was shown that  $\EWG_{\underline2}>\EWG_{\underline3}$
for any choice of parameters (see Appendix~\ref{sec-WG}).
Finally, in case (iv), Cubitus repressor protein is not present in either the second or fourth cells:
\beq
   \begin{array}{ll}
    \CI_{1}=U_{1}\frac{1}{1+H_{\tCI}C_{\tCI}}, & 
      \CN_{1}=U_1\frac{H_{\tCI}C_{\tCI}}{1+H_{\tCI}C_{\tCI}}\\
    \CI_{2,4}=U_{2,4}, & \CN_{2,4}=0.
   \end{array}
\eeq
Again it is not difficult to see that $\kappa_{\tCI\twg}$, $\kappa_{\tCN\twg}$ may take values
only from components $\GI$ or $\GII$, or $\GA$ (it is always necessary that $U_2>U_4$).

Component $\GA$ will never become connected to any of the other four, due to opposite requirements
on the second cell (compare equations\rf{eq-wgU2} and\rf{eq-wgU2E}).
But the comparison above for $\CI_i$ and $\CN_i$ show that the more general case $T_2\neq T_4$ only
contributes to connect components $\GI$ and $\GII$, all the others remaining disconnected.

\clearpage
\section{Computing the cylindrical algebraic decomposition}
\label{sec-CAD}
A CAD for the parameter space $G$ can be computed from
equations\rf{eq-discrete-hh}-(\ref{eq-discrete-HH}), by imposing the conditions
$\out(x)\in\Y^{\tWT}$, as given by\rf{eq-Y_WT}.
By Theorem~\ref{th-unique-ss}, given $\out(x)$ we can solve for $\EN$, $\EWG$,
$\IWG$, $\PTC$, and $\HH$ uniquely as a function of $\en$, $\wg$, $\ptc$, $\hh$
and the parameters $p$.

First, note that a CAD is not unique, and here we will start by arbitrarily 
chosing the maximal levels for $\wg$, $\ci$, and $\ptc$, that is:
\beqn{eq-CAD1}
  \alpha_{\tCI\twg},\ \alpha_{\tIWG\twg},\ U_1,\ U_2,\ U_4,\ T_2,\ T_1\ \ 
  (\mbox{with } T_1<T_2),
\eeqn
with physiological constraints as listed in Table~\ref{par-free}.
A hierarchy of conditions can then be computed for the remaining parameters.

Second, note that the parameters appearing on the equations for $\IWG$ and $\EWG$,
as well as those for $\PTC$ and $\HH$, do not appear on any other equation and, 
moreover, the unique solution for these four species has the same form for any set 
of parameters (Theorem~\ref{th-unique-ss}). 
Similarly, all half lives and the Cubitus cleavage rate can also be arbitrarily chosen.
So, we have a second group of parameters which can be arbitrarily chosen, with
no conditions to satisfy except for physiological constraints. 
These parameters are (also listed in Table~\ref{par-free}):
\beqn{eq-CAD2}
  &&   H_{\tWG},\  r_M,\  r_{LM},\  \ren,\  \rex,  \nonumber\\
  &&   H_{\tPTC},\  H_{\tHH},\  [\PTC]_0,\  [\HH]_0,\  
       r_{LM\tPTC},\  r_{LM\tHH},\  \kappa_{\tPTC\tHH}, \\
  &&   H_{\tCI},\ C_{\tCI}.\nonumber
\eeqn
Let $p_{\tfree}$ denote the subfamily of parameters\rf{eq-CAD1} and\rf{eq-CAD2}. 

Third, using the computed unique steady state expressions for $\EN$, $\IWG$, $\EWG$
$\PTC$, $\HH$, $\CI$, and $\CN$ write down the conditions for consistency for the expressions of 
$\en$, $\wg$, $\ptc$, $\ci$, and $\hh$.
We have seen that $\EN$, $\CI$, and $\CN$ have simple expressions:
\beqn{eq-ENwt}
   \EN^{\tWT}=\en^{\tWT}=(0,0,1,0)'
\eeqn
and, from Lemma~\ref{lm-CI-CN},
\beq
  \CI^{\tWT}&=&\frac{1}{1+H_{\tCI}C_{\tCI}}(U_1,U_2,0,U_4), \\
  \CN^{\tWT}&=&\frac{H_{\tCI}C_{\tCI}}{1+H_{\tCI}C_{\tCI}}(U_1,U_2,0,U_4).
\eeq
The steady state expressions for $\IWG$, $\EWG$, $\PTC$, and $\HH$ are more complicated so, 
for simplicity, we will denote them:
\beq
  \EWG&=&\px_{\tEWG}=\px_{\tEWG}(p_{\tfree}),\\
  \IWG&=&\px_{\tIWG}=\px_{\tIWG}(p_{\tfree}),\\
  \PTC&=&\px_{\tPTC}=\px_{\tPTC}(p_{\tfree}),\\
  \HH &=&\px_{\tHH} =\px_{\tHH}(p_{\tfree}).
\eeq
We start by showing that there are other parameters which can be arbitrarily chosen,
and thus complete the proof of Table~\ref{par-free}.
\begin{lemma}\label{lm-par-free}
The parameters $\kappa_{\tEN\tci}$, $\kappa_{\tEN\thh}$, and $\kappa_{\tCN\thh}$
may take arbitrary values in the interval $(0,1)$.
\end{lemma}
\Proof
The requirements for consistency of the $\ci^{\tWT}$ expression are:
\beq
   \EN_1^{\tWT} < \kappa_{\tEN\tci} \ \mbox{ and }\
   \EN_2^{\tWT} < \kappa_{\tEN\tci} \ \mbox{ and }\
   \EN_3^{\tWT} > \kappa_{\tEN\tci} \ \mbox{ and }\
   \EN_4^{\tWT} < \kappa_{\tEN\tci}, 
\eeq
which are clearly satisfied, in view of\rf{eq-ENwt}, for any $\kappa_{\tEN\tci}\in(0,1)$. 
The requirements for consistency of the $\hh^{\tWT}$ expression are:
\beq
  && \EN_1^{\tWT} < \kappa_{\tEN\thh} \ \ \fbox{ or }\ \ 
     U_1\frac{H_{\tCI}C_{\tCI}}{1+H_{\tCI}C_{\tCI}}>\kappa_{\tCN\thh} \\ 
  && \EN_2^{\tWT} < \kappa_{\tEN\thh} \ \ \fbox{ or }\ \ 
     U_2\frac{H_{\tCI}C_{\tCI}}{1+H_{\tCI}C_{\tCI}}> \kappa_{\tCN\thh} \\ 
  && \EN_3^{\tWT} > \kappa_{\tEN\thh}  \ \ \mbox{ and }\ \ 
     0 < \kappa_{\tCN\thh} \\ 
  && \EN_4^{\tWT} < \kappa_{\tEN\thh} \ \ \fbox{ or }\ \ 
     U_4\frac{H_{\tCI}C_{\tCI}}{1+H_{\tCI}C_{\tCI}}> \kappa_{\tCN\thh} 
\eeq
Again in view of\rf{eq-ENwt}, these conditions are all automatically satisfied
for any $\kappa_{\tEN\thh},\kappa_{\tCN\thh}\in(0,1)$.
\qed

\vspace{2mm}
Next, the constraints for the parameters in Table~\ref{par-common} are shown.
\begin{lemma}\label{lm-par-common}
For system\rf{eq-sys} with steady state output set\rf{eq-Y_WT}, the following hold:
\bit
\item[(a)] $\kappa_{\tPTC\tCI}\in(0,\min\{T_1, \PTC_{2,\tT}\} )$; 
\item[(b)] $\kappa_{\tCI\tptc}\in(0,\frac{1}{1+H_{\tCI}C_{\tCI}}\ \min\{U_1,U_2,U_4 \})$;
\item[(c)] $\kappa_{\tCN\tptc}\in(\frac{H_{\tCI}C_{\tCI}}{1+H_{\tCI}C_{\tCI}}\max\{U_1,U_2,U_4 \},1)$;
\item[(d)] Either
      $\kappa_{\tCN\ten}\in(0,\frac{H_{\tCI}C_{\tCI}}{1+H_{\tCI}C_{\tCI}}\min\{U_1,U_2,U_4\})$ 
      and   $\kappa_{\tWG\ten}\in(0,\EWG_{\underline 3})$, \\
 or $\kappa_{\tCN\ten}\in(0,\frac{H_{\tCI}C_{\tCI}}{1+H_{\tCI}C_{\tCI}}\min\{U_1,U_2\})$ 
      and $\kappa_{\tWG\ten}\in(\EWG_{\underline 4},\EWG_{\underline 3})$.   
\eit
\end{lemma}

\Proof
Part (a) follows immediately from Lemma~\ref{lm-CI-CN}, since $\PTC_1^{\tWT}=T_1$ and 
both $\PTC_{1,2}^{\tWT}$ have to be larger than $\kappa_{\tPTC\tCI}$.

To prove parts (b) and (c), consider the requirements for consistency of the $\ptc^{\tWT}$ expression:
\beq
 &&  U_1\frac{1}{1+H_{\tCI}C_{\tCI}} > \kappa_{\tCI\tptc} \ \ \mbox{ and }\ \ 
     U_1\frac{H_{\tCI}C_{\tCI}}{1+H_{\tCI}C_{\tCI}}< \kappa_{\tCN\tptc} \\ 
 &&  U_2\frac{1}{1+H_{\tCI}C_{\tCI}} > \kappa_{\tCI\tptc} \ \ \mbox{ and }\ \ 
     U_2\frac{H_{\tCI}C_{\tCI}}{1+H_{\tCI}C_{\tCI}}< \kappa_{\tCN\tptc} \\
 &&  0 < \kappa_{\tCI\tptc} \ \ \fbox{ or }\ \ 
     0 > \kappa_{\tCN\tptc} \\ 
 &&  U_4\frac{1}{1+H_{\tCI}C_{\tCI}} > \kappa_{\tCI\tptc} \ \ \mbox{ and }\ \ 
     U_4\frac{H_{\tCI}C_{\tCI}}{1+H_{\tCI}C_{\tCI}}< \kappa_{\tCN\tptc}. 
\eeq
The third line is trivially satisfied, while the other lines involve logical ANDs.
These immediately yield conditions (b) and (c).

To prove part (d), consider the requirements for consistency of the $\en^{\tWT}$ expression:
\beq
  && \px_{\tEWG_{\underline 1}} < \kappa_{\tWG\ten} \ \ \fbox{ or }\ \ 
     U_1\frac{H_{\tCI}C_{\tCI}}{1+H_{\tCI}C_{\tCI}}> \kappa_{\tCN\ten} \\ 
  && \px_{\tEWG_{\underline 2}} < \kappa_{\tWG\ten} \ \ \fbox{ or }\ \ 
     U_2\frac{H_{\tCI}C_{\tCI}}{1+H_{\tCI}C_{\tCI}}> \kappa_{\tCN\ten} \\
  && \px_{\tEWG_{\underline 3}} > \kappa_{\tWG\ten} \ \ \mbox{ and }\ \ 
     0 < \kappa_{\tCN\ten} \\ 
  && \px_{\tEWG_{\underline 4}} < \kappa_{\tWG\ten} \ \ \fbox{ or }\ \ 
     U_4\frac{H_{\tCI}C_{\tCI}}{1+H_{\tCI}C_{\tCI}}> \kappa_{\tCN\ten}. 
\eeq
Theorem~\ref{th-EWG-symmetry}, says that 
$\px_{\tEWG_{\underline 4}}<\px_{\tEWG_{\underline 1}}=\px_{\tEWG_{\underline 3}}<
\px_{\tEWG_{\underline 2}}$, so these conditions can be reduced to:
\beqn{eq-en-ka}
  \kappa_{\tWG\ten} < \px_{\tEWG_{\underline 3}}
  \ \ \mbox{ and }\ \ 
  \kappa_{\tCN\ten} < \frac{H_{\tCI}C_{\tCI}}{1+H_{\tCI}C_{\tCI}}\,\min\{U_1,U_2,U_4\},
\eeqn
or
\beqn{eq-en-kb}
  \px_{\tEWG_{\underline 4}} < \kappa_{\tWG\ten} < \px_{\tEWG_{\underline 3}}
  \ \ \mbox{ and }\ \ 
  \kappa_{\tCN\ten} <\frac{H_{\tCI}C_{\tCI}}{1+H_{\tCI}C_{\tCI}}\,\min\{U_1,U_2\}.
\eeqn
It is obvious that the subsets defined by\rf{eq-en-ka} and\rf{eq-en-kb} intersect:
just choose elements 
$\kappa_{\tWG\ten}\in(\px_{\tEWG_{\underline 4}},\px_{\tEWG_{\underline 3}})$ and 
$\kappa_{\tCN\ten} < \frac{H_{\tCI}C_{\tCI}}{1+H_{\tCI}C_{\tCI}}\,\min\{U_1,U_2,U_4\}$.  
\qed

\vspace{2mm}
Lastly, we come to the parameters in Table~\ref{par-regions} and which complete the
characterization of the feasible parameter space.
\begin{theorem}\label{th-regions}
The set $G$ consists of five disconnected regions of parameters:
\beq
   G=\GI\cup\GII\cup\GIII\cup\GIV\cup\GA
\eeq
each of the regions characterized by Tables~\ref{par-free},~\ref{par-common},~\ref{par-regions}.
\end{theorem}

\Proof
The only parameters whose possible intervals have not yet been found are
$\kappa_{\tWG\twg}$, $\kappa_{\tCI\twg}$, and $\kappa_{\tCN\twg}$.
Consider now the requirements for consistency of the $\wg^{\tWT}$ expression.
There are three distinct cases, depending on wether activation of $\wg$ is 
autocatalytic, or through the $\CI$ pathway, or through both.
\bit
\item[] Case 1: both $\CI$ and $\IWG$ contribute to activation of $\wg$.
Here $\wg_2^{\tWT}=\frac{\alpha_{\tCI\twg}+\alpha_{\tWG\twg}}{1+\alpha_{\tCI\twg}+\alpha_{\tWG\twg}}$, 
so set $\px_{\tIWG}=\px_{\tIWG}^{\tCI,\tWG}$.
\beq
    \left(
     U_1\frac{1}{1+H_{\tCI}C_{\tCI}} < \kappa_{\tCI\twg} \ \ \fbox{ or }\ \ 
     U_1\frac{H_{\tCI}C_{\tCI}}{1+H_{\tCI}C_{\tCI}}> \kappa_{\tCN\twg} 
     \right)
     &\mbox{ and }& \px_{\tIWG_1}^{\tCI,\tWG} < \kappa_{\tWG\twg}\\
     \left(
     U_2\frac{1}{1+H_{\tCI}C_{\tCI}} > \kappa_{\tCI\twg} \ \ \mbox{ and }\ \ 
     U_2\frac{H_{\tCI}C_{\tCI}}{1+H_{\tCI}C_{\tCI}}< \kappa_{\tCN\twg}
     \right)
     &\mbox{ and }& \px_{\tIWG_2}^{\tCI,\tWG} > \kappa_{\tWG\twg}\\
     \left(
     0 < \kappa_{\tCI\twg} \ \ \fbox{ or }\ \ 
     0 > \kappa_{\tCN\twg} 
     \right)
     &\mbox{ and }& \px_{\tIWG_3}^{\tCI,\tWG} < \kappa_{\tWG\twg}\\
     \left(
     U_4\frac{1}{1+H_{\tCI}C_{\tCI}} < \kappa_{\tCI\twg} \ \ \fbox{ or }\ \ 
     U_4\frac{H_{\tCI}C_{\tCI}}{1+H_{\tCI}C_{\tCI}}> \kappa_{\tCN\twg} 
     \right)
     &\mbox{ and }& \px_{\tIWG_4}^{\tCI,\tWG} < \kappa_{\tWG\twg}.\\
\eeq

\item[] Case 2: only $\CI$ contributes to activation of $\wg$
Here $\wg_2^{\tWT}=\frac{\alpha_{\tCI\twg}}{1+\alpha_{\tCI\twg}}$, 
so set $\px_{\tIWG}=\px_{\tIWG}^{\tCI}$.
\beq
    \left(
     U_1\frac{1}{1+H_{\tCI}C_{\tCI}} < \kappa_{\tCI\twg} \ \ \fbox{ or }\ \ 
     U_1\frac{H_{\tCI}C_{\tCI}}{1+H_{\tCI}C_{\tCI}}> \kappa_{\tCN\twg} 
     \right)
     &\mbox{ and }& \px_{\tIWG_1}^{\tCI} < \kappa_{\tWG\twg}\\
     \left(
     U_2\frac{1}{1+H_{\tCI}C_{\tCI}} > \kappa_{\tCI\twg} \ \ \mbox{ and }\ \ 
     U_2\frac{H_{\tCI}C_{\tCI}}{1+H_{\tCI}C_{\tCI}}< \kappa_{\tCN\twg} 
     \right)
     &\mbox{ and }& \px_{\tIWG_2}^{\tCI} < \kappa_{\tWG\twg}\\
     \left(
     0 < \kappa_{\tCI\twg} \ \ \fbox{ or }\ \ 
     0 > \kappa_{\tCN\twg} 
     \right)
     &\mbox{ and }& \px_{\tIWG_3}^{\tCI} < \kappa_{\tWG\twg}\\
     \left(
     U_4\frac{1}{1+H_{\tCI}C_{\tCI}} < \kappa_{\tCI\twg} \ \ \fbox{ or }\ \ 
     U_4\frac{H_{\tCI}C_{\tCI}}{1+H_{\tCI}C_{\tCI}}> \kappa_{\tCN\twg} 
     \right)
     &\mbox{ and }& \px_{\tIWG_4}^{\tCI} < \kappa_{\tWG\twg}.\\
\eeq

\item[] Case 3: only $\IWG$ contributes to activation of $\wg$.
Here $\wg_2^{\tWT}=\frac{\alpha_{\tWG\twg}}{1+\alpha_{\tWG\twg}}$, 
so set $\px_{\tIWG}=\px_{\tIWG}^{\tWG}$.
\beq
     \left(
     U_1\frac{1}{1+H_{\tCI}C_{\tCI}} < \kappa_{\tCI\twg} \ \ \fbox{ or }\ \ 
     U_1\frac{H_{\tCI}C_{\tCI}}{1+H_{\tCI}C_{\tCI}}> \kappa_{\tCN\twg} 
     \right)
     &\mbox{ and }& \px_{\tIWG_1}^{\tWG} < \kappa_{\tWG\twg}\\
     \left(
     U_2\frac{1}{1+H_{\tCI}C_{\tCI}} < \kappa_{\tCI\twg} \ \ \fbox{ or }\ \ 
     U_2\frac{H_{\tCI}C_{\tCI}}{1+H_{\tCI}C_{\tCI}}> \kappa_{\tCN\twg} 
     \right)
      &\mbox{ and }& \px_{\tIWG_2}^{\tWG} > \kappa_{\tWG\twg}\\
     \left(
     0 < \kappa_{\tCI\twg} \ \ \fbox{ or }\ \ 
     0 > \kappa_{\tCN\twg} 
     \right)
     &\mbox{ and }& \px_{\tIWG_3}^{\tWG} < \kappa_{\tWG\twg}\\
     \left(
     U_4\frac{1}{1+H_{\tCI}C_{\tCI}} < \kappa_{\tCI\twg} \ \ \fbox{ or }\ \ 
     U_4\frac{H_{\tCI}C_{\tCI}}{1+H_{\tCI}C_{\tCI}}> \kappa_{\tCN\twg} 
     \right)
     &\mbox{ and }& \px_{\tIWG_4}^{\tWG} < \kappa_{\tWG\twg}.\\
\eeq
\eit
Note that a set of parameters that satisfies case 3 cannot satisfy any of the other 
two. This is because of the conditions on $U_2$. For cases 1 and 2:
\beq
     U_2\frac{1}{1+H_{\tCI}C_{\tCI}} > \kappa_{\tCI\twg} \ \ \mbox{ and }\ \ 
     U_2\frac{H_{\tCI}C_{\tCI}}{1+H_{\tCI}C_{\tCI}}< \kappa_{\tCN\twg}, 
\eeq 
while in case 3, the condition to be satisfied is exactly the negation of this.  
In other words, the region of parameter space defined by case 3 {\it cannot} be 
connected to regions defined by cases 1 and 2.

In addition, note that in cases 1 and 2, $U_2=U_4$ leads to empty intervals for $\kappa_{\tCI\twg}$, 
$\kappa_{\tCN\twg}$). And a similarly conclusion holds when $U_2=U_1$. 
This leads to four disconnected regions defined by:
$U_1>U_2>U_4$, $U_2>U_1,U_4$, $U_2<U_1,U_4$ and $U_1<U_2<U_4$. 
It is now easy to check that, in each of these five regions, the intervals for the three parameters 
$\kappa_{\tWG\twg}$, $\kappa_{\tCI\twg}$, and $\kappa_{\tCN\twg}$ are as given in  
Table~\ref{par-regions}.
\qed

\begin{table}[ht]
\caption{Free parameters (physiological constraints, as in~\cite{dmmo00}).}
\label{par-free}
\begin{center}
\begin{tabular}{lc}
\hline 
  Parameter  & Interval \\
\hline
  $U_{1,2,4}$, $T_2$    & $(0,1]$ \\
  $T_1$    & $(0,T_2)$ \\
  $\alpha_{\tCI\twg}$,  $\alpha_{\tWG\twg}$ & $(0,1)$\\ 
   & \\
\hline
  Half-lives   & $[5,100]$ \\
   ($H_X$)      & \\  
   & \\
\hline
  Transfer, cleavage & $(0,1)$ \\
  ($C_{\tCI}$, $\rex$, $\ren$, $r_M$, $r_{LM}$, & \\
   $r_{LM\tPTC}$, $r_{M\tHH}$, $[\HH]_0$, $[\PTC]_0$, $\kappa_{\tPTC\tHH}$) & \\
   & \\
\hline
  $\kappa_{\tEN\tci}$, $\kappa_{\tEN\thh}$, $\kappa_{\tCN\thh}$& $(0,1)$ \\
   & \\
\hline
\end{tabular}
\end{center}
\end{table}

\begin{table}[th]
\caption{Parameters with constraints common to all regions.}
\label{par-common}
\begin{center}
\begin{tabular}{lc}
\hline
  Parameter  & Interval \\
\hline
  $\kappa_{\tPTC\tCI}$ & $(0,\min\{T_1, \px_{\tPTC_{2,\tT}}\} )$ \\
    & \\
\hline
  $\kappa_{\tCI\tptc}$ & $(0,\frac{1}{1+H_{\tCI}C_{\tCI}}\ \min\{U_1,U_2,U_4 \})$\\
    &  \\
\hline
  $\kappa_{\tCN\tptc}$ & $(\frac{H_{\tCI}C_{\tCI}}{1+H_{\tCI}C_{\tCI}}\max\{U_1,U_2,U_4 \},1)$\\
    &  \\
\hline
    & $\kappa_{\tCN\ten}\in(0,\frac{H_{\tCI}C_{\tCI}}{1+H_{\tCI}C_{\tCI}}\min\{U_1,U_2,U_4\})$ 
      and   $\kappa_{\tWG\ten}\in(0,\px_{\tEWG_{\underline 3}})$ \\
  $\kappa_{\tCN\ten}$, $\kappa_{\tWG\ten}$      & or \\ 
    & $\kappa_{\tCN\ten}\in(0,\frac{H_{\tCI}C_{\tCI}}{1+H_{\tCI}C_{\tCI}}\min\{U_1,U_2\})$ 
      and $\kappa_{\tWG\ten}\in(\px_{\tEWG_{\underline 4}},\px_{\tEWG_{\underline 3}})$   \\
\hline
\end{tabular}
\end{center}

\end{table}
\begin{table}[h]
\caption{The five disconnected components.}
\label{par-regions}
\begin{center}
\begin{tabular}{lccc}
\hline
  Parameter  & Interval & Region \\
\hline
  $\kappa_{\tCI\twg}$  & $(\max\{U_1,U_4\}\frac{1}{1+H_{\tCI}C_{\tCI}},
                          U_2\frac{1}{1+H_{\tCI}C_{\tCI}})$
                       &  \\
    & & $\GI$  \\
  $\kappa_{\tCN\twg}$  & $(U_2\frac{H_{\tCI}C_{\tCI}}{1+H_{\tCI}C_{\tCI}},1)$
                       & $U_2>U_4,U_1$\\
   & &  \\
\hline
  $\kappa_{\tCI\twg}$  & $(U_4\frac{1}{1+H_{\tCI}C_{\tCI}},
                           U_2\frac{1}{1+H_{\tCI}C_{\tCI}})$
                       &   \\
    & & $\GII$  \\
  $\kappa_{\tCN\twg}$  & $(U_2\frac{H_{\tCI}C_{\tCI}}{1+H_{\tCI}C_{\tCI}},
                           U_1\frac{H_{\tCI}C_{\tCI}}{1+H_{\tCI}C_{\tCI}})$
                       & $U_1>U_2>U_4$ \\
   & &  \\
\hline
  $\kappa_{\tCI\twg}$  & $(0,U_2\frac{1}{1+H_{\tCI}C_{\tCI}})$
                       &  \\
    & & $\GIII$   \\
  $\kappa_{\tCN\twg}$  & $(U_2\frac{H_{\tCI}C_{\tCI}}{1+H_{\tCI}C_{\tCI}},
                           \min\{U_1,U_4\}\frac{H_{\tCI}C_{\tCI}}{1+H_{\tCI}C_{\tCI}})$
                       & $U_1,U_4>U_2$  \\
   & & \\
\hline
  $\kappa_{\tCI\twg}$  & $(U_1\frac{1}{1+H_{\tCI}C_{\tCI}},
                           U_2\frac{1}{1+H_{\tCI}C_{\tCI}})$
                       &  \\
    & & $\GIV$  \\
  $\kappa_{\tCN\twg}$  & $(U_2\frac{H_{\tCI}C_{\tCI}}{1+H_{\tCI}C_{\tCI}},
                           U_4\frac{H_{\tCI}C_{\tCI}}{1+H_{\tCI}C_{\tCI}})$
                       & $U_4>U_2>U_1$  \\
   & &  \\
\hline
  & $\kappa_{\tCI\twg}\in(U_i\frac{1}{1+H_{\tCI}C_{\tCI}},1)$ or
    $\kappa_{\tCN\twg}\in(0,U_i\frac{H_{\tCI}C_{\tCI}}{1+H_{\tCI}C_{\tCI}})$ & \\
 $\kappa_{\tCI\twg}$, $\kappa_{\tCN\twg}$  & & $\GA$ \\
  & for all $i=1,2,4$ & \\
   & & \\
\hline
                       & $(\max\{\px_{\tIWG_{1,2,3,4}}^{\tCI}\}, 1)$ &  \\
   $\kappa_{\tWG\twg}$ & or & $\GI$,$\GII$,$\GIII$,$\GIV$  \\
                       & $(\max\{\px_{\tIWG_{1,3,4}}^{\tCI,\tWG}\},\px_{\tIWG_{2}}^{\tCI,\tWG} )$ 
                       &  \\
   & &  \\
\hline
   $\kappa_{\tWG\twg}$ & $(\max\{\px_{\tIWG_{1,3,4}}^{\tWG}\},\px_{\tIWG_{2}}^{\tWG} )$ 
                       & $\GA$ \\
   & & \\
\hline
\end{tabular}
\end{center}
\end{table}

\begin{figure}
\centerline{
\scalebox{0.25}[0.25]{\includegraphics{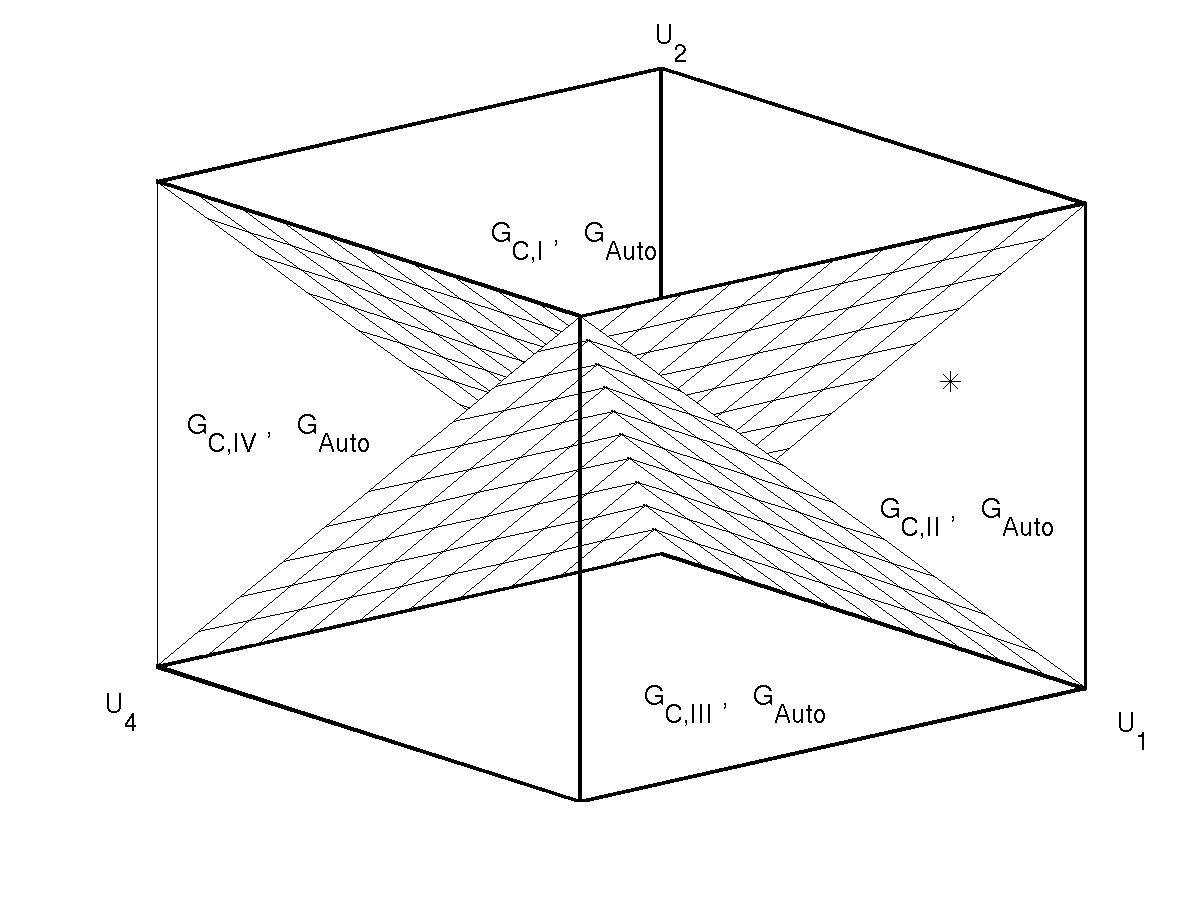} } }
\caption{Projection of set $G$ into the $(U_1,U_2,U_4)$ space.
The regions are defined by the planes $U_2=U_4$ and $U_2=U_1$.
In region $\GA$, $U_i$ can take values in the whole unitary cube, while $\GI$ through $\GIV$ 
do not include any of the points in the two planes. In this figure $\GA$ appears to ``intersect''
all others, since they share values of $U_i$. However, this is only the projection effect, since
not all parameters can be shown.}
\label{fig-cubeU124}
\end{figure}

\begin{figure}
\centerline{
\scalebox{0.25}[0.25]{\includegraphics{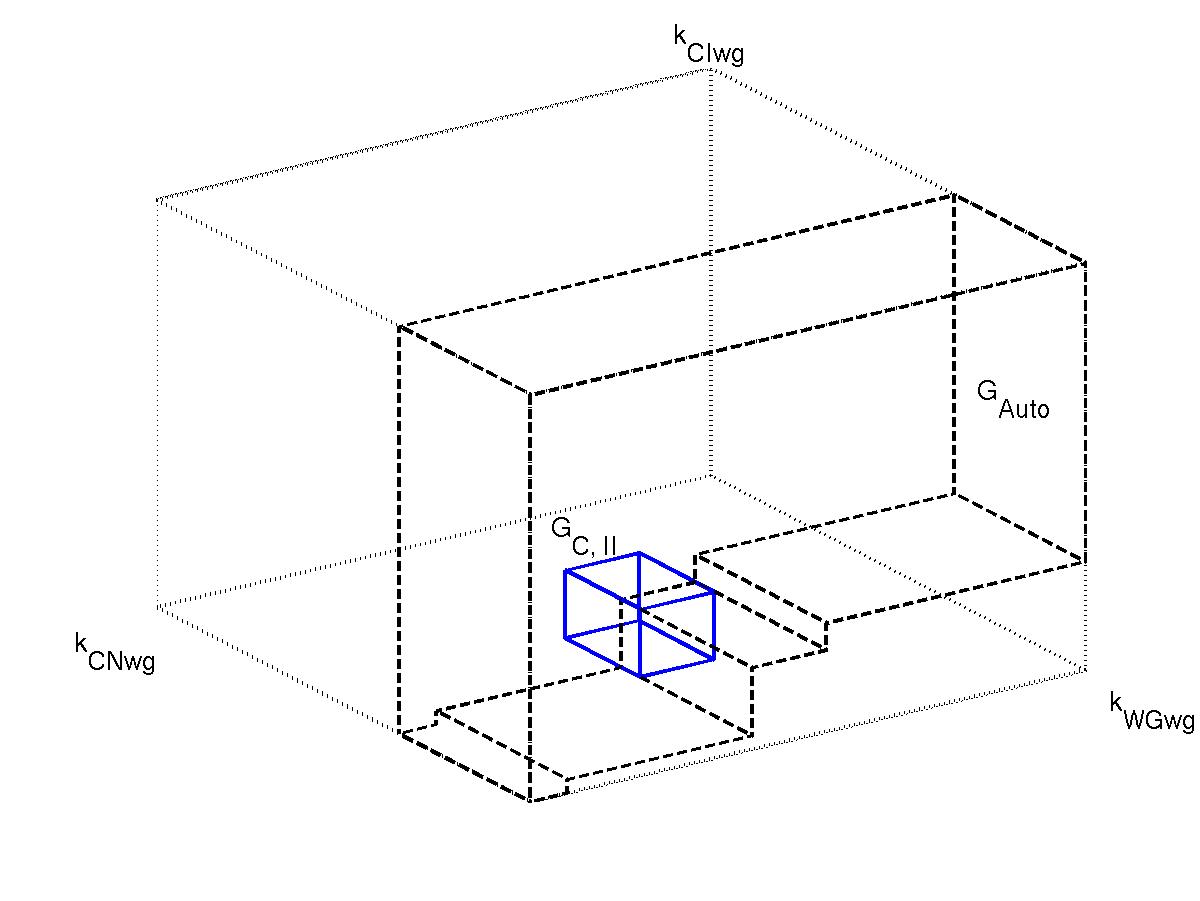} } }
\caption{An example of regions $\GII$ (solid line rectangle) and $\GA$ (dashed line polyhedron). 
This is the projection on the space $(\kappa_{\tCI\twg},(\kappa_{\tCN\twg},(\kappa_{\tWG\twg})$,
of the fibre over the point represented by ``$*$'' in Fig.~\ref{fig-cubeU124}.
This points corresponds to choosing values for $(U_1,U_2,U_4)$ in region $\GII$.}
\label{fig-cube-kwg}
\end{figure}

\begin{figure}
\centerline{
\scalebox{0.5}[0.5]{\includegraphics{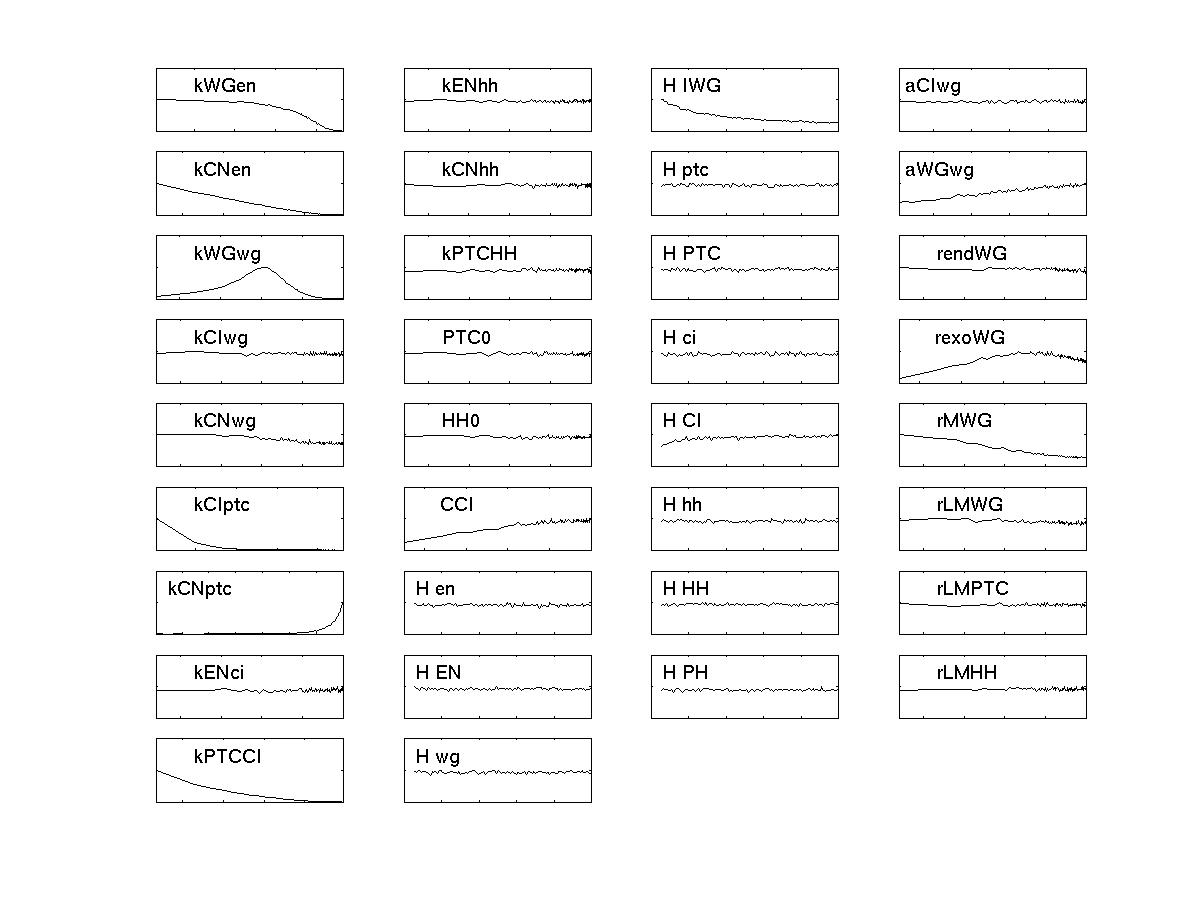} } }
\caption{Parameter histograms out of 70026 parameter sets (refer to model equations for 
explanation of parameters). The notation and scales follow those of Fig. 6 in~\cite{do02}. 
The half-lifes (denoted Hx) range between 5 and 100 mins in a linear scale.
The coefficients aCIwg and aWGwg range between 1.0 and 10.0 also in a linear scale.
All other parameters range between $10^{-3}$ and 1, in log$_{10}$ scale. }
\label{fig-par-dist}
\end{figure}

\begin{figure}
\centerline{
\scalebox{0.45}[0.45]{\includegraphics{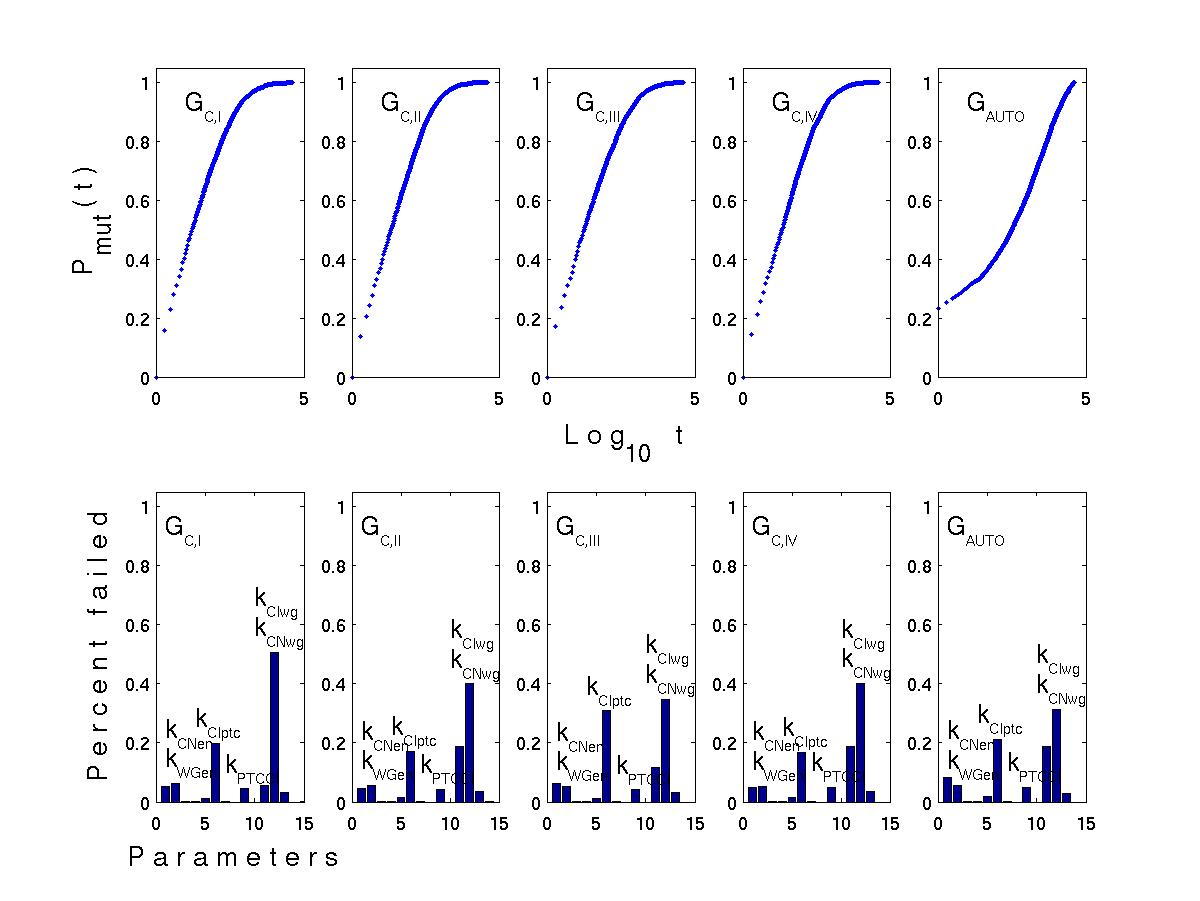} } }
\caption{The distribution functions for the probability of leaving the region (or ``mutation''), 
in each of the four regions. $P_{\mut}(t)$ is shown in the top row, where the $x$ axis is in
logarithmic scale, and $t$ ranges from $0$ to $40000$ (40000 is the maximal number of steps 
allowed in the random walks).
The bottom row shows the failed parameters and the percentage of cases where each parameter failed 
(exit through a certain ``face'' of the polygonal regions).}
\label{fig-prob-mutation}
\end{figure}

\end{document}